\documentclass[english,10pt,aps,prd,a4paper,preprintnumbers,floatfix,nofootinbib,showpacs,superscriptaddress,notitlepage,hyperref]{revtex4-2} 
\usepackage{amssymb}
\usepackage{slashed}
\usepackage{amsmath}
\usepackage{mathtools}
\usepackage{braket}
\usepackage{float}
\usepackage[utf8]{inputenc}
\usepackage{graphicx}
\usepackage{subfigure}
\usepackage{latexsym}
\usepackage{rotating}
\usepackage{epic}
\usepackage{eepic}
\usepackage{epsfig}
\usepackage{color}
\usepackage{cancel}
\usepackage{natbib}
\usepackage[colorlinks=true,citecolor=darkred,urlcolor=darkred, pdfborder={0 0 0}]{hyperref}
\usepackage[normalem]{ulem}

\usepackage[height=8.85in,width=6.4in]{geometry}

\usepackage[height=8.85in,width=6.4in]{geometry}

\newcommand{\bea}{\begin{eqnarray}}
\newcommand{\eea}{\end{eqnarray}}

\newcommand\dd{\mathrm{d}}

\definecolor{darkred}{rgb}{0.6,0,0}

\definecolor{linkcolor}{rgb}{0,0,0.5}

\begin{document}

\title{Chiral $Z^\prime$ in FASER, FASER2, DUNE, and ILC beam dump experiments}

\author{Kento Asai}
\email{kento@icrr.u-tokyo.ac.jp}
\affiliation{Institute for Cosmic Ray Research (ICRR), The University of Tokyo, Kashiwa, Chiba 277--8582, Japan}
\affiliation{Department of Physics, Faculty of Engineering Science, Yokohama National University, Yokohama 240--8501, Japan}
\author{Arindam Das}
\email{arindamdas@oia.hokudai.ac.jp}
\affiliation{Institute for the Advancement of Higher Education, Hokkaido University, Sapporo 060-0817, Japan}
\affiliation{Department of Physics, Hokkaido University, Sapporo 060-0810, Japan}
\author{Jinmian Li}
\email{jmli@scu.edu.cn}
\affiliation{College of Physics, Sichuan University, Chengdu 610065, China}
\author{Takaaki Nomura}
\email{nomura@scu.edu.cn}
\affiliation{College of Physics, Sichuan University, Chengdu 610065, China}
\author{Osamu Seto}
\email{seto@particle.sci.hokudai.ac.jp}
\affiliation{Department of Physics, Hokkaido University, Sapporo 060-0810, Japan}
\vskip .4in
\bibliographystyle{unsrt} 
\begin{abstract}\noindent
The origin of neutrino mass is a big unsolved problem of the Standard Model (SM) that motivate us to consider beyond the SM (BSM) scenarios where SM-singlet right-handed neutrinos (RHNs) are introduced to explain the origin of the light neutrino masses through the seesaw mechanism. 
There is a variety of ways which could lead us to this goal and one of them is a general U$(1)$ extension of the SM. 
In this scenario, three SM-singlet RHNs are introduced to cancel the gauge and mixed gauge gravity anomalies. 
After anomaly cancellation, we notice that the left- and right-handed charged fermions are differently charged under the general U$(1)$ gauge group evolving a chiral scenario. 
After the breaking of the general U$(1)$ symmetry, a neutral BSM gauge boson $(Z^\prime)$ acquires mass and it is a free parameter.
Such $Z^\prime$, being lighter than $5$ GeV, could be probed at the intensity and lifetime frontiers like FASER, FASER2, DUNE, and ILC beam dump experiments. 
The estimated bounds are needed to be compared with the existing bounds. 
We find that existing constraints from Orsay, Nomad, PS191, KEK, LSND, CHARM experiments, and cosmological scenario like SN1987A can be compared in our case once estimated for chiral scenarios. 
Finally, we compare the parameter spaces showing viable ones that could be probed by FASER, FASER2, DUNE, and ILC beam dump experiments and already excluded regions from Orsay, Nomad, PS191, KEK, LSND, CHARM, and SN1987A for a chiral scenario.
\end{abstract}
\vspace{-3cm}
\maketitle

\preprint{EPHOU-22-010}

\setcounter{page}{1}
\setcounter{footnote}{0}
\section{Introduction}
\label{sec:introduction}
Observations of the tiny neutrino mass and flavor mixing \cite{ParticleDataGroup:2020ssz} by different terrestrial experiments can not be explained within the framework of the Standard Model (SM). 
Studies of galaxy rotation curve, bullet cluster, and large-scale cosmological data indicate that nearly 25$\%$ of the energy budget of the Universe is allotted for the nonluminous objects called dark matter~\cite{Planck:2018vyg}, which is another aspect for thinking beyond the SM (BSM). 
To accommodate neutrino mass, a simple but interesting aspect called the seesaw mechanism was proposed where the light neutrino mass can be explained by the suppression in terms of a large mass scale where the SM is extended by SM-singlet heavy right-handed neutrinos (RHNs)~\cite{Minkowski:1977sc,Yanagida:1979as,Gell-Mann:1979vob,Mohapatra:1979ia,Schechter:1980gr}. 
Such a scenario is an appropriate realization of the Weinberg operator where a dimension five operator can be predicted with the SM where a heavy mass scale can be integrated out and lepton number can be violated by a unit of 2~\cite{Weinberg:1979sa}. 
The implementation of the seesaw mechanism is possible in the context of a U$(1)$ gauge extension of the SM, namely B$-$L (baryon minus lepton), where three generations of right-handed neutrinos are incorporated to acquire an anomaly free model ~\cite{Davidson:1979wr,Mohapatra:1980qe,Mohapatra:1980qe,Marshak:1979fm,Davidson:1978pm,Davidson:1987mh,Wetterich:1981bx,Masiero:1982fi,Mohapatra:1982xz,Buchmuller:1991ce}. 
After the B$-$L symmetry breaking, the Majorana mass terms of the heavy neutrinos are generated which further help to achieve tiny neutrino masses through the seesaw mechanism. 
In addition to that, several attempts are made to explain the origin of the observed DM relic abundance predicting a variety of BSM scenarios \cite{Bertone:2004pz,Arcadi:2017kky}.  

There is another interesting scenario where the SM gauge group~:~$\mathcal{G}_{\rm SM} = {\rm SU}(3)_C \otimes {\rm SU}(2)_L \otimes {\rm U}(1)_Y$ can be extended by a general U$(1)_X$ gauge group. 
To cancel the gauge and mixed gauge-gravity anomalies, we introduce three generations of the right-handed neutrinos. 
After breaking of the general U$(1)_X$ symmetry, the Majorana mass term of the heavy neutrino is generated which helps to explain the origin of light neutrino mass after the electroweak symmetry breaking~\cite{Das:2016zue,Das:2017flq,Das:2017deo}. 
Apart from the neutrino mass, such a scenario has an additional aspect giving rise to a BSM and neutral gauge boson called $Z^\prime$ which acquires mass term after the general U$(1)_X$ symmetry breaking. 
Like the heavy neutrinos, the mass of $Z^\prime$ is a free parameter. 
Unlike the $B-L$ scenario, the general U$(1)_X$ extension has an interesting property. 
After the anomaly cancellation, we find that the U$(1)_X$ charges of the left- and right-handed SM fermions are different. 
As a result, they interact differently with the $Z^\prime$ invoking chiral interactions. 
In the context of high energy electron positron collider, the chiral property of the heavy $Z^\prime$ has been tested in Ref.~\cite{Das:2021esm-3}. 
These models have been well studied in the context of different phenomenological aspect; however, the implication of the chiral nature of the $Z^\prime$ was unexplored in the context of intensity and lifetime frontiers. 
As a result, we study these chiral scenarios to estimate the constraints on the model parameters from FASER, FASER2, DUNE, and ILC beam dump experiments systematically for the first time.

In this paper, we study two different general U$(1)_X$ extensions of the SM. 
The first case consists of three generations of the RHNs which are charged equally under the general U$(1)_X$ gauge group, and three generations that participate in the neutrino mass generation mechanism through the seesaw mechanism. 
On the other hand, in the second case, two of the three generations of the RHNs are equally charged under the general U$(1)_X$ gauge group, but the third one is differently charged \cite{Das:2017flq}. 
In the second case, we have two Higgs doublets and three SM-singlet scalars which are differently charged under the general U$(1)_X$ gauge group. 
Because of the general U$(1)_X$ gauge symmetry one of the Higgs doublets participates in the neutrino mass generation mechanism whereas the other one has Yukawa interactions with the SM fermions making the model neutrinophilic \cite{Das:2017deo}. 

The $Z^\prime$ mass being a free parameter could be light $(\leq 5~\rm GeV)$, and it can be tested at different low and high energy experiments of the lifetime and intensity frontiers. 
To study such a scenario, we formally implement our chiral model setup into the context of different beam dump experiments~\cite{Bauer:2018onh,Ilten:2018crw} such as proton beam dump in the context of DUNE~\cite{Dev:2021qjj}, electron/ positron beam dump in that of International Linear Collider (ILC)~\cite{Asai:2020xnz,Asai:2021ehn}, and the Frascati PADME experiment~\cite{Nardi:2018cxi}, and FASER (ForwArd Search ExpeRiment)~\cite{Feng:2017uoz,FASER:2018eoc,Araki:2020wkq}, which is a new experiment to search for long-lived particles coming from the ATLAS interaction point, where B$-$L, $L_i-L_j$~\cite{Foot:1990mn,He:1990pn,He:1991qd,Heeck:2011wj,Altmannshofer:2014cfa,Chun:2018ibr,Han:2019zkz} and dark photon scenarios have been proposed to probe successfully at different masses of the neutral BSM gauge boson respectively. 
As we mentioned above, the $Z^\prime$ interacts differently with the left-handed and right-handed charged fermions in our model setup.
Therefore, it will affect the search reaches for the $Z^\prime$ boson mass and general U$(1)_X$ gauge coupling depending on the U$(1)_X$ charges of the charged fermions. Finally, due to the chiral nature under our model setup, the $Z^\prime$ will have both vector and axial-vector currents. 

We restrict ourselves for the $Z^\prime$ lighter than 5 GeV for the UV-complete models under consideration. 
In our scenarios the couplings of the SM charged fermions with the $Z^\prime$ do not depend on the generations. 
As a result, we can probe the light $Z^\prime$ $(\leq 0.5~\rm GeV)$ from the supernova muons produced from SN1987A~\cite{Kamiokande-II:1987idp, Bionta:1987qt, Alekseev:1987ej,Burrows:1987zz}. 
Due to the generic U$(1)_X$ coupling between the $Z^\prime$ and charged fermions including muons, we apply this observation to constrain the general U$(1)_X$ scenarios. 
We noticed that a vector-like scenario in the context of U$(1)_{\rm B-L}$ has been studied in Ref.~\cite{Croon:2020lrf}. 
Finally, we compare the bounds from different experiments like Orsay, Nomad, PS191, KEK, LSND, CHARM, and cosmological scenario like SN1987A on the general U$(1)_X$ coupling with $Z^\prime$ mass plane for different U$(1)_X$ charges. 
The existing bounds deal with the vector-like scenarios. 
As a result, we need to estimate the existing bounds for the chiral scenarios. 
Finally, we compare the viable parameter regions obtained from FASER, FASER2, DUNE, and ILC beam dump with the existing bounds from Orsay, Nomad, PS191, KEK, LSND, CHARM, and SN1987A for the chiral scenario.

We organize our article in the following way. The general U$(1)_X$ extensions of the SM studied in this article have been described in Sec.~\ref{sec:model}. 
Methodology of the calculations of the constraints on the general U$(1)_X$ gauge coupling for different $Z^\prime$ masses is shown in Sec.~\ref{sec:calc}.
We discuss the results obtained in Sec.~\ref{sec:summary}. 
Finally, we conclude our article in Sec.~\ref{sec:conc}.

\section{Model}
\label{sec:model}
We consider a general and gauged U$(1)_X$ extension of the SM involving three generations of RHNs that are required to cancel the gauge and mixed gauge-gravitational anomalies. 
To realize anomaly cancellation, we find that the left- and right-handed fermions are differently charged under the general U$(1)_X$ gauge group. 
It manifests the chiral nature of the $Z^\prime$ interactions with the charged fermions. 
To illustrate such an aspect, we introduce two UV-complete scenarios and explain how the chiral interactions have been evolved.

\subsection{Case-I}
To investigate a general U$(1)_X$ extension of the SM, we introduce a SM-singlet scalar $(\Phi)$ and three generations of the SM-singlet RHNs to cancel gauge and mixed gauge gravity anomalies. 
We write down the particle content of a minimal U$(1)_X$ extension of the SM and the general charges $(x^\prime_f, f= \{ q,~u,~d,~\ell,~e,~\nu \})$ of the particles in Table~\ref{tab1}. 
The U$(1)_X$ charges of the particles are the same for three generations. 
The general charges can be related to each other from the following gauge and mixed gauge-gravity anomaly cancellation conditions:  
\begin{align}
{\rm U}(1)_X \otimes \left[ {\rm SU}(3)_C \right]^2&\ :&
			2x_q^\prime - x_u^\prime - x_d^\prime &\ =\  0~, \nonumber \\
{\rm U}(1)_X \otimes \left[ {\rm SU}(2)_L \right]^2&\ :&
			3x_q^\prime + x_\ell^\prime &\ =\  0~, \nonumber \\
{\rm U}(1)_X \otimes \left[ {\rm U}(1)_Y \right]^2&\ :&
			x_q^\prime - 8x_u^\prime - 2x_d^\prime + 3x_\ell^\prime - 6x_e^\prime &\ =\  0~, \nonumber \\
\left[ {\rm U}(1)_X \right]^2 \otimes {\rm U}(1)_Y &\ :&
			{x_q^\prime}^2 - {2x_u^\prime}^2 + {x_d^\prime}^2 - {x_\ell^\prime}^2 + {x_e^\prime}^2 &\ =\  0~, \nonumber \\
\left[ {\rm U}(1)_X \right]^3&\ :&
			{6x_q^\prime}^3 - {3x_u^\prime}^3 - {3x_d^\prime}^3 + {2x_\ell^\prime}^3 - {x_\nu^\prime}^3 - {x_e^\prime}^3 &\ =\  0~, \nonumber \\
{\rm U}(1)_X \otimes \left[ {\rm grav.} \right]^2&\ :&
			6x_q^\prime - 3x_u^\prime - 3x_d^\prime + 2x_\ell^\prime - x_\nu^\prime - x_e^\prime &\ =\  0~, 
\label{anom-f}
\end{align}
respectively. 
Using the $\mathcal{G}_{\rm SM} \otimes$ U$(1)_X$ gauge symmetry, we write the Yukawa interactions in the following way:
\begin{equation}
{\cal L}^{\rm Yukawa} = - Y_u^{\alpha \beta} \overline{q_L^\alpha} H u_R^\beta
                                - Y_d^{\alpha \beta} \overline{q_L^\alpha} \tilde{H} d_R^\beta
				 - Y_e^{\alpha \beta} \overline{\ell_L^\alpha} \tilde{H} e_R^\beta
				- Y_\nu^{\alpha \beta} \overline{\ell_L^\alpha} H N_R^\beta- Y_N^\alpha \Phi \overline{(N_R^\alpha)^c} N_R^\alpha + {\rm H.c.}~,
\label{LYk}   
\end{equation}
where $H$ is the SM Higgs doublet, and $\tilde{H} = i \tau^2 H^*$ with $\tau^2$ being the second Pauli matrix. 
From the Yukawa interactions, we write the following conditions using U$(1)_X$ neutrality as 
\begin{eqnarray}
-\frac{1}{2} x_H^{} &=& - x_q^\prime + x_u^\prime \ =\  x_q^\prime - x_d^\prime \ =\  x_\ell^\prime - x_e^\prime=\  - x_\ell^\prime + x_\nu^\prime~, \nonumber \\
2 x_\Phi^{}	&=& - 2 x_\nu^\prime~. 
\label{Yuk}
\end{eqnarray} 
Finally, solving Eqs.~\eqref{anom-f} and \eqref{Yuk}, we express the U$(1)_X$ charges of the particles in terms of $x_H^{}$ and $x_\Phi^{}$. 
Hence, we find U$(1)_X$ charges of the particles can be written as a linear combination of the U$(1)_Y$ and B$-$L charges. 
It implies that the left- and right-handed fermions are differently charged under the general U$(1)_X$ gauge group. 
Fixing $x_\Phi^{}=1$ with $x_H^{}=0$, the U$(1)_X$ charges reduce to those of the B$-$L scenario, and with $x_H^{}=-2$, the charges reduce to those of the U$(1)_{\rm R}$ scenario~\cite{Fayet:1990wx,Fayet:2016nyc,Alvarado:2021blq}.
\begin{table}[t]
\begin{center}
\begin{tabular}{||c||ccc||rcr|c||c||c||c||c||c||c||}
\hline
\hline
            & SU(3)$_C^{}$ & SU(2)$_L^{}$ & U(1)$_Y^{}$ & \multicolumn{3}{c|}{U(1)$_X^{}$}&$-2$&$-1$&$-0.5$& $0$& $0.5$ & $1$ & $2$  \\
            &&& &&&&U$(1)_{\rm{R}}$& & &B$-$L&&&  \\
\hline
\hline
&&&&&&&&&&&&&\\[-12pt]
&&&&&&&&&&&&&\\
$q_L^\alpha$    & {\bf 3}   & {\bf 2}& $\frac{1}{6}$ & $x_q^\prime$ 		& = & $\frac{1}{6}x_H^{} + \frac{1}{3}x_\Phi^{}$   &$0$&$\frac{1}{6}$&$\frac{1}{4}$&$\frac{1}{3}$&$\frac{5}{12}$&$\frac{1}{2}$&$\frac{1}{3}$\\
$u_R^\alpha$    & {\bf 3} & {\bf 1}& $\frac{2}{3}$ & $x_u^\prime$ 		& = & $\frac{2}{3}x_H^{} + \frac{1}{3}x_\Phi^{}$   &$-1$&$-\frac{1}{3}$&$0$&$\frac{1}{3}$&$\frac{1}{2}$&$1$&$\frac{5}{3}$\\
$d_R^\alpha$    & {\bf 3} & {\bf 1}& $-\frac{1}{3}$ & $x_d^\prime$ 		& = & $-\frac{1}{3}x_H^{} + \frac{1}{3}x_\Phi^{}$  &$1$&$\frac{2}{3}$&$\frac{1}{2}$&$\frac{1}{3}$&$\frac{1}{6}$&$0$&$-\frac{1}{3}$\\
\hline
\hline
&&&&&&&&&&&&&\\
$\ell_L^\alpha$    & {\bf 1} & {\bf 2}& $-\frac{1}{2}$ & $x_\ell^\prime$ 	& = & $- \frac{1}{2}x_H^{} - x_\Phi^{}$   &$0$&$-\frac{1}{2}$&$-\frac{3}{4}$&$-1$&$\frac{5}{4}$&$-\frac{3}{2}$&$-2$ \\
$e_R^\alpha$   & {\bf 1} & {\bf 1}& $-1$   & $x_e^\prime$ 		& = & $- x_H^{} - x_\Phi^{}$   &$1$&$0$&$-\frac{1}{2}$&$-1$&$-\frac{3}{2}$&$-2$&$-3$ \\
\hline
\hline
&&&&&&&&&&&&&\\
$N_R^\alpha$   & {\bf 1} & {\bf 1}& $0$   & $x_\nu^\prime$ 	& = & $- x_\Phi^{}$  &$-1$&$-1$&$-1$&$-1$&$-1$&$-1$&$-1$ \\
\hline
\hline
&&&&&&&&&&&&&\\
$H$         & {\bf 1} & {\bf 2}& $-\frac{1}{2}$  &  $-\frac{1}{2}x_H^{}$ 	& = & $-\frac{1}{2}x_H^{}$\hspace*{7.0mm} &$1$&$\frac{1}{2}$&$\frac{1}{4}$&$0$&$-\frac{1}{4}$&$-\frac{1}{2}$&$-1$ \\ 
$\Phi$      & {\bf 1} & {\bf 1}& $0$  &  $2 x_\Phi^{}$ 	& = & $2 x_\Phi^{}$ &$2$&$2$&$2$&$2$&$2$&$2$&$2$  \\ 
\hline
\hline
\end{tabular}
\end{center}
\caption{
Particle content of the minimal U$(1)_X$ model with general U$(1)_X$ charges before and after anomaly cancellation, and $\alpha$ stands for three generations of the fermions. Considering different benchmark values of the $x_H^{}$ fixing $x_\Phi^{}=1$, we estimate different charges of the left- and right-handed fermions of the U$(1)_X$ model. Here, $x_H^{}=0$ is the B$-$L case, which is a purely vector-like scenario, studied in this article as a reference.}
\label{tab1}
\end{table}
Similarly, with $x_H^{}=-1$, the U$(1)_X$ charge of the right-handed electron $(e_R^{})$ is zero, with $x_H^{}=-0.5$, the U$(1)_X$ charge of the right-handed up-type quark $(u_R^{})$ is zero, and with $x_H^{}=1$, the U$(1)_X$ charge of the right-handed down-type quark $(d_R^{})$ is zero, respectively. 
For the other benchmark points like $x_H^{}=0.5$ and $2$ considered in Table~\ref{tab1}, the U$(1)_X$ charges of all the SM fermions are nonzero. 
In this model, we consider $g_X^{}$ as the U$(1)_X$ gauge coupling which is a free parameter appearing as either $g^\prime x_H^{}$ or $g^\prime x_\Phi^{}$ in the interaction Lagrangian. 
In the case of a general U$(1)_X$ extension of the SM, there is a neutral BSM gauge boson $Z^\prime$ which interacts with particles in the model. 
After the anomaly cancellation, we notice that the left- and right-handed fermions interact differently with the $Z^\prime$ leading to the chiral nature of the neutral gauge boson that will be investigated in the article further.

The renormalizable Higgs potential of this model is given by
\begin{align}
  V \ = \ m_H^2(H^\dag H) + \lambda_H^{} (H^\dag H)^2 + m_\Phi^2 (\Phi^\dag \Phi) + \lambda_\Phi^{} (\Phi^\dag \Phi)^2 + \lambda_{\rm mix} (H^\dag H)(\Phi^\dag \Phi)~,
\end{align}
where $H$ and $\Phi$ are the SM Higgs doublet and SM-singlet scalar, respectively, which can be approximated separately in the analysis of scalar potential where $\lambda_{\rm mix}$ is very small \cite{Oda:2015gna,Das:2016zue}. 
After U$(1)_X^{}$ gauge symmetry and electroweak symmetry are broken, the scalar fields $H$ and $\Phi$ potentially develop their vacuum expectation values (VEVs) as 
\begin{align}
  \braket{H} \ = \ \frac{1}{\sqrt{2}}\begin{pmatrix} v+h\\0 
  \end{pmatrix}~, \quad {\rm and}\quad 
  \braket{\Phi} \ =\  \frac{v_\Phi^{}+\phi}{\sqrt{2}}~,
\end{align}
where electroweak scale is marked with $v=246$ GeV at the potential minimum and $v_\Phi^{}$ is considered to be a free parameter. 
After the U$(1)_X$ symmetry breaking with a limit $v_\Phi^{} \gg v$, the mass term of the neutral BSM gauge boson $Z^\prime$ is evolved and can be defined as 
\begin{equation}
 m_{Z^\prime}^{}=  2 g_X^{}  v_\Phi^{}~,
\end{equation}
for $x_\Phi^{}=1$. The $Z^\prime$ mass is a free parameter in this model, and $g_X^{}$ is the U$(1)_X$ coupling. 
From Eq.~\eqref{LYk}, we find that the RHNs interact with $\Phi$, which can generates the Majorana mass term for the heavy neutrinos after the U$(1)_X$ symmetry breaking. 
After the electroweak symmetry breaking, the Dirac Yukawa mass term is generated which finally induces the seesaw mechanism to explain the origin of tiny neutrino mass term and flavor mixing. 
The Majorana and Dirac mass terms can be written as 
\begin{equation}
    m_{N_R^\alpha}^{} \ = \ \frac{Y^\alpha_{N}}{\sqrt{2}} v_\Phi^{}, \, \, \, \, \,
    m_{D}^{\alpha \beta} \  =  \ \frac{Y_{\nu}^{\alpha \beta}}{\sqrt{2}} v~,
\label{mDI}
\end{equation}
respectively. 
The neutrino mass mixing can be written as
\begin{equation}
   m_\nu= \begin{pmatrix} 0&m_D^{}\\m_D^T&m_N^{} \end{pmatrix}~,
\label{num-1}
\end{equation}
and the light neutrino mass eigenvalues $\sim -m_D^{} m_N^{-1} m_D^T$ can be obtained by diagonalizing Eq.~\eqref{num-1}. 
In this paper, the neutrino mass generation is not the main subject. 
Therefore, we are not investigating the properties of the light and heavy neutrinos in this article. 
The mass of the $Z^\prime$ and the U$(1)_X$ gauge coupling are constrained by the previous studies of LEP~\cite{LEP:2003aa}, Tevatron~\cite{Carena:2004xs}, and LHC~\cite{Amrith:2018yfb} from different decay modes~\cite{Das:2021esm-3}.

\subsection{Case-II}
\begin{table}[t]
\begin{center}
\begin{tabular}{||c||ccc||rcl|c||c||c||c||c||c||c||}
\hline
\hline
            & SU(3)$_C$ & SU(2)$_L$ & U(1)$_Y$ & \multicolumn{3}{c|}{U(1)$_X$}&$-2$&$-1$&$-0.5$& $0$& $0.5$ & $1$ & $2$  \\
            &&& &&&&U$(1)_{\rm{R}}$& & &B$-$L&&&  \\
\hline
\hline
&&&&&&&&&&&&&\\[-12pt]
&&&&&&&&&&&&&\\
$q_L^\alpha$    & {\bf 3}   & {\bf 2}& $\frac{1}{6}$ & $x_q^\prime$ 		& = & $\frac{1}{6}x_H^{} + \frac{1}{3}$   &$0$&$\frac{1}{6}$&$\frac{1}{4}$&$\frac{1}{3}$&$\frac{5}{12}$&$\frac{1}{2}$&$\frac{1}{3}$\\
$u_R^\alpha$    & {\bf 3} & {\bf 1}& $\frac{2}{3}$ & $x_u^\prime$ 		& = & $\frac{2}{3}x_H^{} + \frac{1}{3}$   &$-1$&$-\frac{1}{3}$&$0$&$\frac{1}{3}$&$\frac{1}{2}$&$1$&$\frac{5}{3}$\\
$d_R^\alpha$    & {\bf 3} & {\bf 1}& $-\frac{1}{3}$ & $x_d^\prime$ 		& = & $-\frac{1}{3}x_H^{} + \frac{1}{3}$  &$1$&$\frac{2}{3}$&$\frac{1}{2}$&$\frac{1}{3}$&$\frac{1}{6}$&$0$&$-\frac{1}{3}$\\
\hline
\hline
&&&&&&&&&&&&&\\
$\ell_L^\alpha$    & {\bf 1} & {\bf 2}& $-\frac{1}{2}$ & $x_\ell^\prime$ 	& = & $- \frac{1}{2}x_H^{} - 1$   &$0$&$-\frac{1}{2}$&$-\frac{3}{4}$&$-1$&$\frac{5}{4}$&$-\frac{3}{2}$&$-2$ \\
$e_R^\alpha$   & {\bf 1} & {\bf 1}& $-1$   & $x_e^\prime$ 		& = & $- x_H^{} - 1$   &$1$&$0$&$-\frac{1}{2}$&$-1$&$-\frac{3}{2}$&$-2$&$-3$ \\
\hline
\hline
&&&&&&&&&&&&&\\
$N_R^{1,2}$   & {\bf 1} & {\bf 1}& $0$   & $x_\nu^\prime$ 	& = & $-4$  &$-4$&$-4$&$-4$&$-4$&$-4$&$-4$&$-4$ \\
$N_R^3$   & {\bf 1} & {\bf 1}& $0$   & $x_\nu^{\prime\prime}$ 	& = & $5$  &$5$&$5$&$5$&$5$&$5$&$5$&$5$ \\
\hline
\hline
&&&&&&&&&&&&&\\
$H_1$         & {\bf 1} & {\bf 2}& $-\frac{1}{2}$  &  $x_{H_{1}}^{}$ 	& = & $-\frac{x_H^{}}{2}$ &$1$&$\frac{1}{2}$&$\frac{1}{4}$&$0$&$-\frac{1}{4}$&$-\frac{1}{2}$&$-1$ \\ 
$H_2$         & {\bf 1} & {\bf 2}& $-\frac{1}{2}$  &  $x_{H_{2}}^{}$ 	& = &  $-\frac{1}{2} x_{H}^{}+3$ &$4$&$\frac{7}{2}$&$\frac{13}{2}$&$3$&$\frac{11}{4}$&$\frac{5}{2}$&$2$ \\ 
$\Phi_1$      & {\bf 1} & {\bf 1}& $0$  &  $ x_{\Phi_{1}}^{}$ 	& = & $+8$ &$+8$&$+8$&$+8$&$+8$&$+8$&$+8$&$+8$  \\ 
$\Phi_2$      & {\bf 1} & {\bf 1}& $0$  &  $x_{\Phi_{2}}^{}$ 	& = & $-10$ &$-10$&$-10$&$-10$&$-10$&$-10$&$-10$&$-10$  \\ 
$\Phi_3$      & {\bf 1} & {\bf 1}& $0$  &  $x_{\Phi_{3}}^{}$ 	& = & $-3$ &$-3$&$-3$&$-3$&$-3$&$-3$&$-3$&$-3$  \\ 
\hline
\hline
\end{tabular}
\end{center}
\caption{
Particle content of the alternative U$(1)_X$ model with general U$(1)_X$ charges before and after anomaly cancellation, and $\alpha = 1,2,3$ stands for three generations of the fermions. Considering different benchmark values of the $x_H^{}$, we estimate different charges of the left- and right-handed fermions of the U$(1)_X$ model. Here, $x_H^{}=0$ is an alternative B$-$L case, which is a purely vector-like scenario, studied in this article as a reference.}
\label{tab2}
\end{table}

There is another interesting U(1)$_X$ extension of the SM whose minimal particle content is shown in Table~\ref{tab2}. 
We call it an alternative U$(1)_X$ scenario. 
Using Eq.~\eqref{anom-f}, we cancel the gauge and mixed gauge-gravity anomalies due to the presence of the general U$(1)_X$ scenario. 
The U$(1)_X$ charge $x_H^{}$ is a real parameter that allows us to manifest the chiral nature of the $Z^\prime$ interactions with the left- and right-handed fermions. 
In this scenario, the three generations of the RHNs are differently charged under the U$(1)_X$. 
The first two generations have U$(1)_X$ charge $-4$, and the third one's charge is $+5$ being a unique choice to cancel the gauge and mixed gauge-gravity anomalies~\cite{Montero:2007cd}.

There are two Higgs doublets $(H_1, H_2)$ that have been introduced in this model. 
In addition to that, there are three SM-singlet scalars $(\Phi_{1,2,3})$. 
The Higgs doublet $H_2$, a second Higgs doublet, only interacts with the lepton doublet and the first two generations of the RHNs due to the U$(1)_X$ charge assignment. 
As a result, it generates the Dirac mass term for the first two generations of the RHNs with U$(1)_X$ charge $-4$. 
The Yukawa couplings of the remaining RHN $(N_R^3)$ with the Higgs doublets are forbidden due to the U$(1)_X$ charges. 
In addition to that, the Majorana mass terms of $N_R^{1,2}$ are generated due to the Majorana Yukawa coupling with SM-singlet scalar $\Phi_{1}$ after the U$(1)_X$ symmetry breaking. 
That for $N_R^3$ is also generated from the VEV of $\Phi_2$. As there is no Dirac mass term obtained involving $N_R^3$ due to the preservation of U$(1)_X$ symmetry, $N_R^3$ does not participate in the tree-level neutrino mass generation mechanism. As a result, the interaction Lagrangian of the RHNs is given by
\bea
-\mathcal{L} _{\rm int}& \ \supset \ & \sum_{\alpha=1}^{3} \sum_{\beta=1}^{2} Y_{1}^{\alpha \beta} \overline{\ell_L^\alpha} H_2 N_R^\beta+\frac{1}{2} \sum_{\alpha=1}^{2} Y_{2}^{\alpha}  \Phi_1 \overline{(N_R^\alpha)^{c}} N_R^\alpha
+\frac{1}{2} Y_{3} \Phi_2 \overline{(N_R^3)^{c}} N_R^3+ \rm{H. c.}~,
\label{ExoticYukawa}
\eea 
where without the loss of generality, we assume $Y_2$ is diagonal. The potential of the scalar fields involved in the model can be written as  
\bea
  V&\ =\ &
m_{H_1}^2 (H_1^\dagger H_1) + \lambda_{H_1}  (H_1^\dagger H_1)^2 + m_{H_2}^2 (H_2^\dagger H_2) + \lambda_{H_2}  (H_2^\dagger H_2)^2 \nonumber \\
&& + m_{\Phi_1}^2 (\Phi_1^\dagger \Phi_1) + \lambda_1  (\Phi_1^\dagger \Phi_1)^2 
+ m_{\Phi_2}^2 (\Phi_2^\dagger \Phi_2) + \lambda_2   (\Phi_2^\dagger \Phi_2)^2 \nonumber \\
&&+ m_{\Phi_3}^2 (\Phi_3^\dagger \Phi_3) + \lambda_3   (\Phi_3^\dagger \Phi_3)^2 
+ ( \mu \Phi_3 (H_1^\dagger H_2) + {\rm H.c.} )  \nonumber \\
&&+ \lambda_4 (H_1^\dagger H_1) (H_2^\dagger H_2)+ \lambda_5 (H_1^\dagger H_2) (H_2^\dagger H_1) +\lambda_6 (H_1^\dagger H_1) (\Phi_1^\dagger \Phi_1)\nonumber \\
&&+ \lambda_7 (H_1^\dagger H_1) (\Phi_2^\dagger \Phi_2)+ \lambda_8 (H_1^\dagger H_2) (\Phi_3^\dagger \Phi_3) +\lambda_9 (H_2^\dagger H_2) (\Phi_1^\dagger \Phi_1)  \nonumber \\
&&+ \lambda_{10} (H_1^\dagger H_1) (\Phi_2^\dagger \Phi_2)+ \lambda_{11} (H_1^\dagger H_2) (\Phi_3^\dagger \Phi_3)+  \lambda_{12} (\Phi_1^\dagger \Phi_1) (\Phi_2^\dagger \Phi_2) \nonumber \\
&&+ \lambda_{13} (\Phi_2^\dagger \Phi_2) (\Phi_3^\dagger \Phi_3)+ \lambda_{14} (\Phi_3^\dagger \Phi_3) (\Phi_1^\dagger \Phi_1)~.
\label{HiggsPotential-2}
\eea
Choosing suitable parameters for the scalar fields in the model to develop their respective VEVs, we write 
\bea
  \braket{H_1} \ = \  \frac{1}{\sqrt 2}\left(  \begin{array}{c}  
    v_{h_1} \\
    0 \end{array}
\right)~,   \; 
  \braket{H_2} \ = \   \frac{1}{\sqrt{2}} \left(  \begin{array}{c}  
    v_{h_2}\\
    0 \end{array}
\right)~,  
  \braket{\Phi_1} \ = \  \frac{v_{1}}{\sqrt{2}}~,  \; 
  \braket{\Phi_2} \ = \  \frac{v_{2}}{\sqrt{2}}~,  \; 
  \braket{\Phi_3} \ = \  \frac{v_{3}}{\sqrt{2}}~,~~~~ 
\eea   
with the condition, $v_{h_1}^2 + v_{h_2}^2 = (246 \,  {\rm GeV})^2$. 

In this model setup, we consider negligibly small quartic couplings between the SU$(2)_L$ doublets and SM-singlet scalars ensuring higher order mixing between the RHNs after the U$(1)_X$ breaking to be strongly suppressed. 
Thanks to the proposed gauge symmetry, the doublet and singlet scalar sectors interact only through the triple coupling $\Phi_3 (H_1^\dagger H_2)+{\rm H.c.}$ which has no significant effect on determining the VEVs $(v_{1,2,3})$ of the singlet scalars $(\Phi_{1,2,3})$ due to the collider constraints $v_1^2 + v_2^2+ v_3^2 \gg v_{h_1}^2 + v_{h_2}^2$. 
As a result, we arrange the parameters in the scalar potential in such a way that the acquired VEVs of the SM-singlet scalars will be almost the same and $\mu < v_1$ respectively. 
Hence, we comment that the third SM-singlet scalar $\Phi_3$ can be considered as a spurion field to generate the mixing between the two Higgs doublets $H_{1,2}$ through the $\mu$ term in Eq.~\eqref{HiggsPotential-2}. 
After $\Phi_3$ acquires VEV such as $\braket{\Phi_3}=v_3/\sqrt{2}$, the mixing between $H_{1,2}$ could be expressed as $m_{\rm mix}^2=\mu v_2/\sqrt{2}$ allowing the part of the potential of the Higgs doublet sector resembling that of the two Higgs doublet model. 
The U$(1)_X$ symmetry ensures that there is no mixing term present among $\Phi_{1,2,3}$. 
Hence, in our model, there are two existing physical Nambu-Goldstone (NG) bosons, which are originated from the SM-singlet scalars, and they are phenomenologically harmless because they are decoupled from the SM thermal bath in the early Universe by the tiny scalar quartic couplings and gauge coupling. 
We consider the SM-singlet scalars to be heavier than $Z^\prime$ so that $Z^\prime$ will not decay into the NG bosons. 
Following U$(1)_X$ symmetry breaking the $Z^\prime$ boson mass could be given as 
\bea
 m_{Z^\prime} = g_X \sqrt{64 v_{1}^2+ 100 v_{2}^2+ 9v_3^2 +\frac{1}{4} x_H^2 v_{h_{1}}^2 + \left(-\frac{1}{2} x_H +3\right)^2  v_{h_{2}}^2}
\simeq g_X \sqrt{64 v_{1}^2+ 100 v_{2}^2+ 9 v_{3}^2}~.
\label{masses-Alt}   
\eea 
The $Z^\prime$ mass is a free parameter of the model and $g_X$ is the U$(1)_X$ coupling.

The Majorana masses of the heavy neutrinos are generated after the U$(1)_X$ symmetry breaking and are written as
\bea
 m_{N_R^{1,2}}=\frac{Y_2^{1,2}}{\sqrt{2}} v_1,\;\; m_{N_R^3} = \frac{Y_3^{3}}{\sqrt{2}} v_2~,
\label{mN3II} 
\eea
using the collider constraints to set $(v_1^2 + v_2^2+ v_3^2) \gg (v_{h_{1}}^2 + v_{h_{2}}^2)$, and  
the Dirac mass terms for the first two generations of neutrinos are generated by $\braket{H_2}$, and it is given by 
\bea
m_{D}^{\alpha \beta} \ = \ \frac{Y_{1}^{\alpha \beta}}{\sqrt{2}} \, v_{h_{2}}~, \label{mDII}
\eea
which further participates in the seesaw mechanism to generate tiny neutrino masses and flavor mixing. 
The neutrino mass matrix can be written as
\bea
m_\nu= \begin{pmatrix} 0&m_D^{}\\m_D^T&m_N^{} \end{pmatrix}~,
\label{num}
\eea
and the light neutrino mass eigenvalues $\sim -m_D^{} m_N^{-1} m_D^T$ can be obtained by diagonalizing Eq.~\eqref{num}.
We have already mentioned that due to the U$(1)_X$ charges, only $N_R^{1,2}$ are involved in the minimal seesaw mechanism~\cite{Smirnov:1993af,King:1999mb,Frampton:2002qc,Ibarra:2003up} while $N_R^3$ has neither direct involvement in the neutrino mass generation nor any interaction with the SM sector allowing $N_R^3$ to be considered as a potential dark matter (DM) candidate.
Due to the U$(1)_X$ symmetry, the first Higgs doublet $(H_1)$ does not couple with the RHNs and the neutrino Dirac masses are generated by the VEV of the second Higgs doublet $(H_2)$ according to Eq.~\eqref{mDII}. 
As a result, this setup can be considered as a sort of the neutrinophilic scenario in the two Higgs doublet model (2HDM) framework~\cite{Ma:2000cc,Wang:2006jy,Gabriel:2006ns,Davidson:2009ha,Haba:2010zi}. 
In Eq.~\eqref{HiggsPotential-2}, we may consider $0 < m_{\rm mix}^2 = \mu v_3/\sqrt{2} \ll m_{\Phi_3}^2$ which further leads to $v_{h_{2}} \sim m_{\rm mix}^2 v_{h_{1}}/m_{\Phi_{3}}^2 \ll v_{h_{1}}$~\cite{Ma:2000cc}.

\subsection{$Z^\prime$ interactions with the fermions}
From the above two examples, we find that charged fermions interact with $Z^\prime$ in the same way if we consider $x_\Phi^{}=1$ in case-I. 
Due to the presence of the general U$(1)_X$ charges, the interaction Lagrangian between the $Z^\prime$ boson and the SM fermions $(f)$ including the quarks and leptons can be written as
\bea
\mathcal{L}^{f} = -g_X (\overline{f}\gamma_\mu q_{f_{L}^{}}^{} P_L^{} f+ \overline{f}\gamma_\mu q_{f_{R}^{}}^{}  P_R^{} f) Z_\mu^\prime~,
\label{Lag1}
\eea
where $P_{L(R)}^{}= (1 \pm \gamma_5)/2$. 
The quantities $q_{f_{L}^{}}^{}$ and $q_{f_{R}^{}}^{}$ are the U$(1)_X$ charges of the left- and right-handed fermions of the cases-I and II. 
From Eq.~\eqref{Lag1}, we write the vector and axial-vector couplings $(C_{V, A})$ for the charged and neutral fermions of our models in Table~\ref{tab-3} considering $x_\Phi^{}=1$ in case-I and that will reproduce the couplings of the fermions with $Z^\prime$ in case-II. 
In this case, the charges of the fermions of case-II will be the same as those of case-I for different values of $x_H$ resulting in the same couplings with $Z^\prime$.
\begin{table}[t]
\begin{center}
\begin{tabular}{|c|c|c|}
\hline\hline
      Type of fermion  & Vector coupling $(C_V)$& Axial-vector coupling $(C_A)$  \\ 
                       & $\frac{q_{f_L}^{}+q_{f_R}^{}}{2}$& $\frac{q_{f_L}^{}-q_{f_R}^{}}{2}$\\
\hline\hline
Charged lepton $(\ell^\alpha)$ &$-\frac{3}{4}x_H^{}-1$&$\frac{1}{4} x_H^{}$\\
SM-like neutrinos $(\nu_L^\alpha)$&$\frac{1}{4} x_H^{}+\frac{1}{2}$&$\frac{1}{4} x_H^{}+\frac{1}{2}$\\
\hline
\hline
Up-type quarks $(q_{u^\alpha})$&$\frac{5}{12} x_H^{}+\frac{1}{3}$&$-\frac{1}{4} x_H^{}$\\
Down-type quarks $(q_{d^\alpha})$&$-\frac{1}{12} x_H^{}+\frac{1}{3}$&$-\frac{1}{4} x_H^{}$\\
\hline\hline
\end{tabular}
\end{center}
\caption{Vector and axial-vector couplings of different SM fermions with $Z^\prime$ involved in the general U$(1)_X$ scenarios and $\alpha$ stands for the three generations of the fermions. The axial-vector couplings for the charged fermions vanish for the B$-$L case. 
}
\label{tab-3}
\end{table}  
Hence, the interactions between $Z^\prime$ and the SM fermions depend on the choices of $x_H^{}$ and $x_\Phi^{}$ manifesting the chiral nature of the model. 
Using Eq.~\eqref{Lag1}, we calculate the partial decay widths of $Z^\prime$ into different SM fermions. 
The partial decay width of the $Z^\prime$ into a pair of a single generation of charged fermions can be written as 
\begin{align}
\label{eq:width-ll}
    \Gamma(Z' \to \bar{f} f)
    &= N_C^{} \frac{m_{Z'}^{} g_{X}^2}{24 \pi} \left[ \left( q_{f_L^{}}^2 + q_{f_R^{}}^2 \right) \left( 1 - \frac{m_f^2}{m_{Z'}^2} \right) + 6 q_{f_L^{}}^{} q_{f_R^{}}^{} \frac{m_f^2}{m_{Z'}^2} \right]~,
\end{align}    
where $m_f$ is the SM fermion mass where $N_C^{}=1$ for the SM leptons and $3$ for the SM quarks.
The partial decay width of the $Z^\prime$ boson into a pair of light neutrinos for three generations can be written as 
\begin{align}   
\label{eq:width-nunu}
    \Gamma(Z' \to \nu \nu)
    = 3 \frac{m_{Z'}^{} g_{X}^2}{24 \pi} q_{f_L^{}}^2~,
\end{align} 
neglecting the tiny neutrino mass. Here $q_{f_L^{}}^{}$ is the U$(1)_X$ charge of the SM lepton doublet. 
The $Z^\prime$ gauge boson can decay into a pair of heavy Majorana neutrinos and the corresponding partial decay width in one generation of the RHN pair can be written as
\begin{align}
\label{eq:width-NN}
    \Gamma(Z' \to N_R^\alpha N_R^\alpha)
    = \frac{m_{Z'}^{} g_{X}^2}{24 \pi} q_{N_R^{}}^2 \left( 1 - \frac{M_N^2}{m_{Z'}^2} \right)^{\frac{3}{2}}~,
\end{align}
with $q_{N_R^{}}^{}$ being the U$(1)_X$ charge of the RHNs, and $m_N^{}$ is the RHN mass. From case-I, we find that $q_{N_R^{}}^{}=x_\Phi^{}=1$ and from case-II we find that $q_{N_R^{}}^{}=-4$ or $5$ depending on the generation of the RHN. In this analysis, we consider that the masses of RHNs are larger than half of the $Z^\prime$ mass not allowing the $Z^\prime \to N_R^{}N_R^{}$ mode. Hence, the total decay width of $Z^\prime$ in this setup can be given by
\begin{align}
    \Gamma_{\rm total}
    = \Gamma(Z' \to {\rm hadrons}) + \Gamma(Z' \to \bar{\nu}\nu)+ \sum_{\ell=e,\mu,\tau} \Gamma(Z' \to \bar{\ell}\ell)~.
\end{align}
Hence, we can estimate the branching ratio for the visible mode as 
\begin{align}
 \rm{BR}( Z^\prime \to \rm{visible})= 1- \frac{\Gamma(Z^\prime \to 2\nu)}{\Gamma_{\rm total}}~,  
\end{align}
where the word ``visible'' stands for all the possible decay modes of $Z^\prime$ except neutrinos.

\begin{figure}[h]
\includegraphics[width=0.495\textwidth]{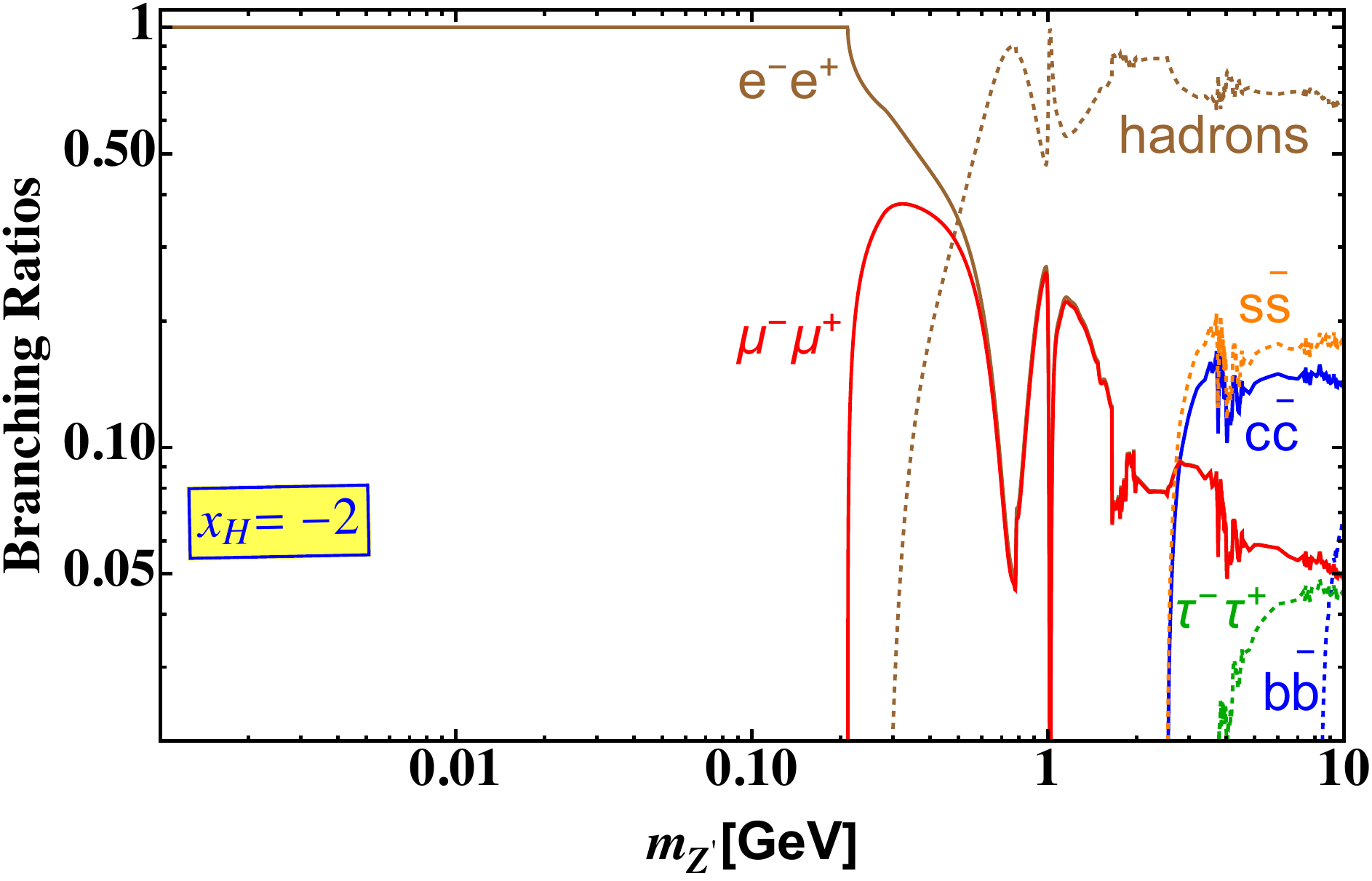}
\includegraphics[width=0.495\textwidth]{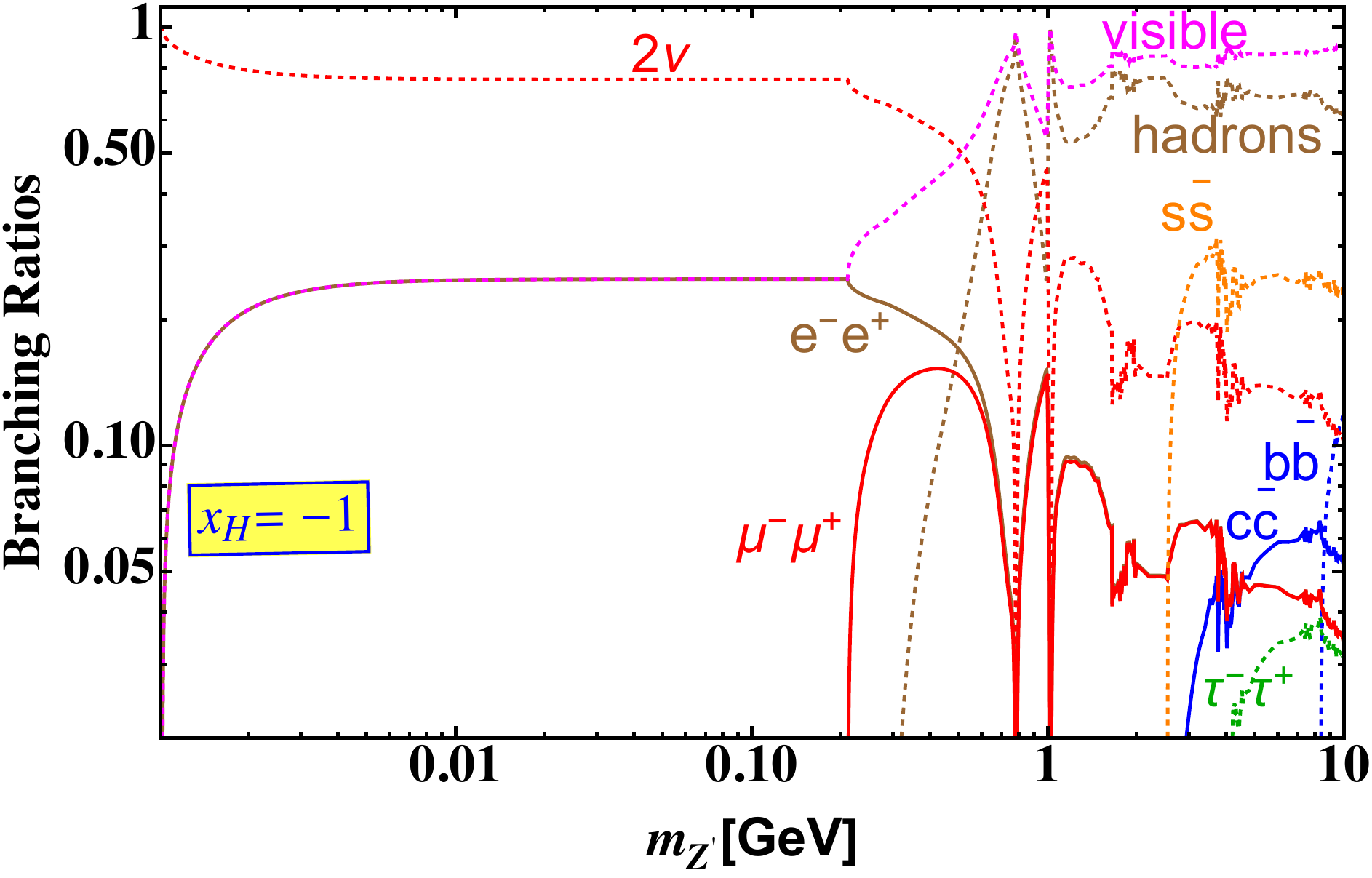}
\includegraphics[width=0.495\textwidth]{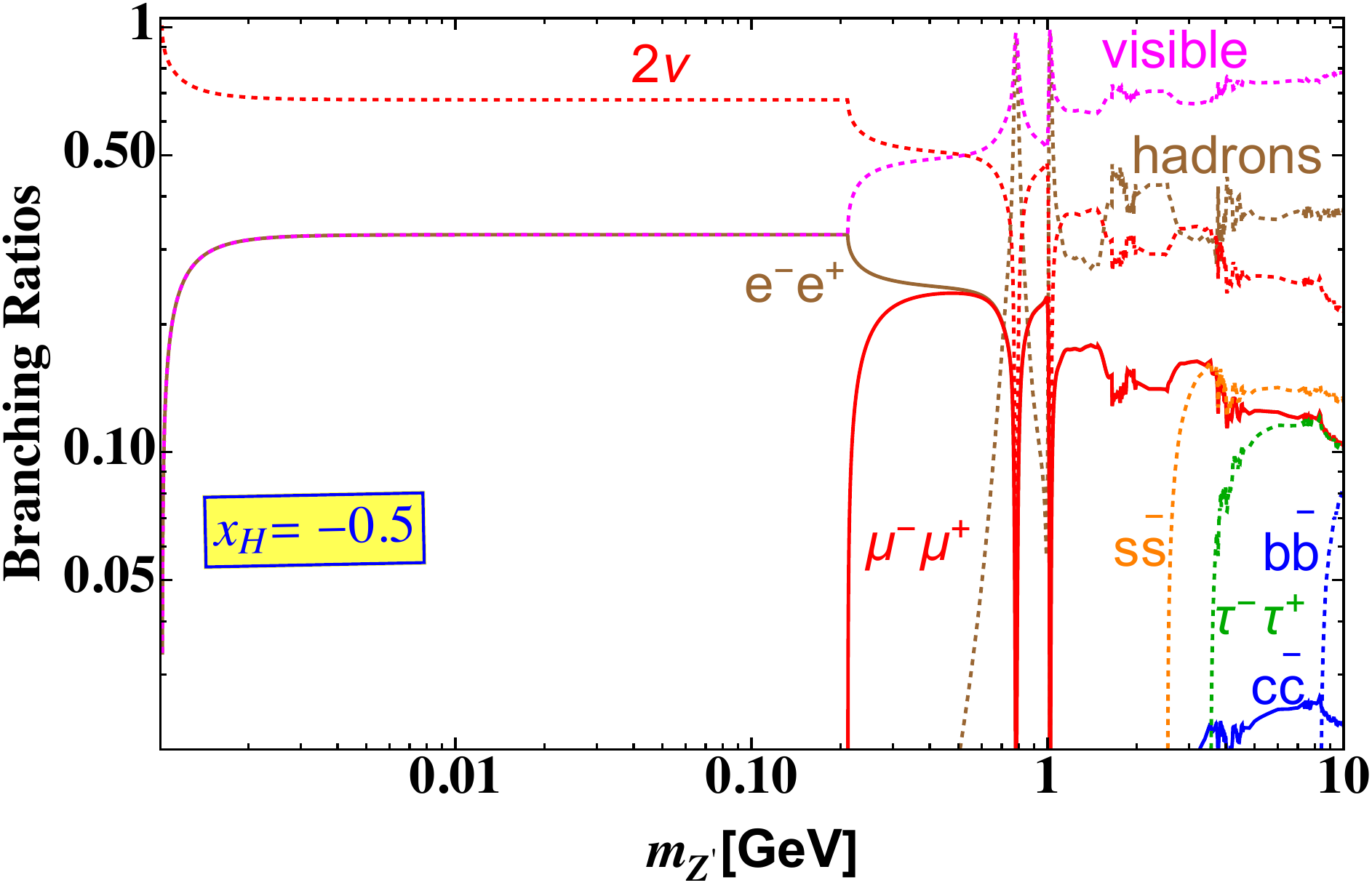}
\includegraphics[width=0.495\textwidth]{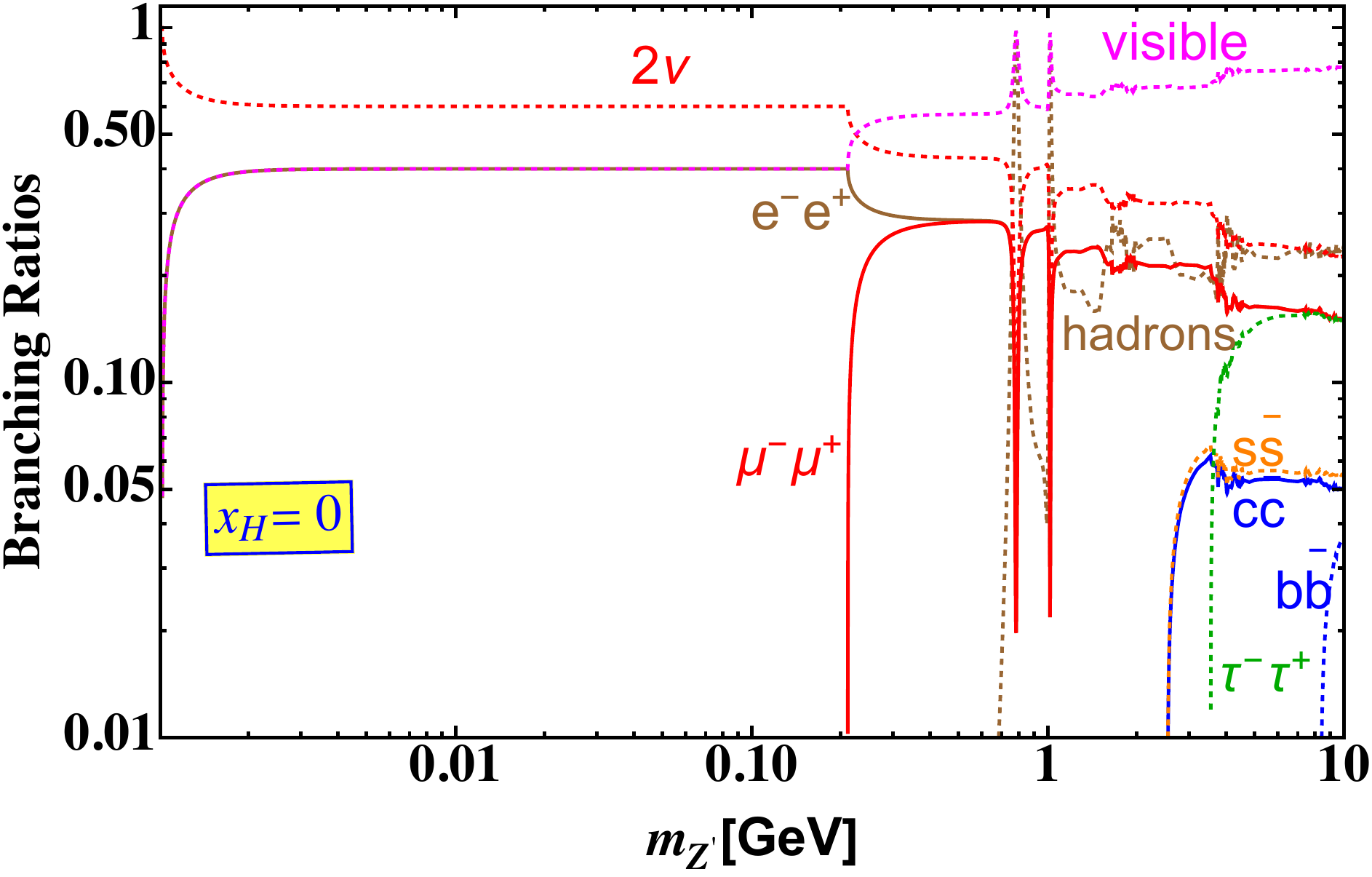}
\includegraphics[width=0.495\textwidth]{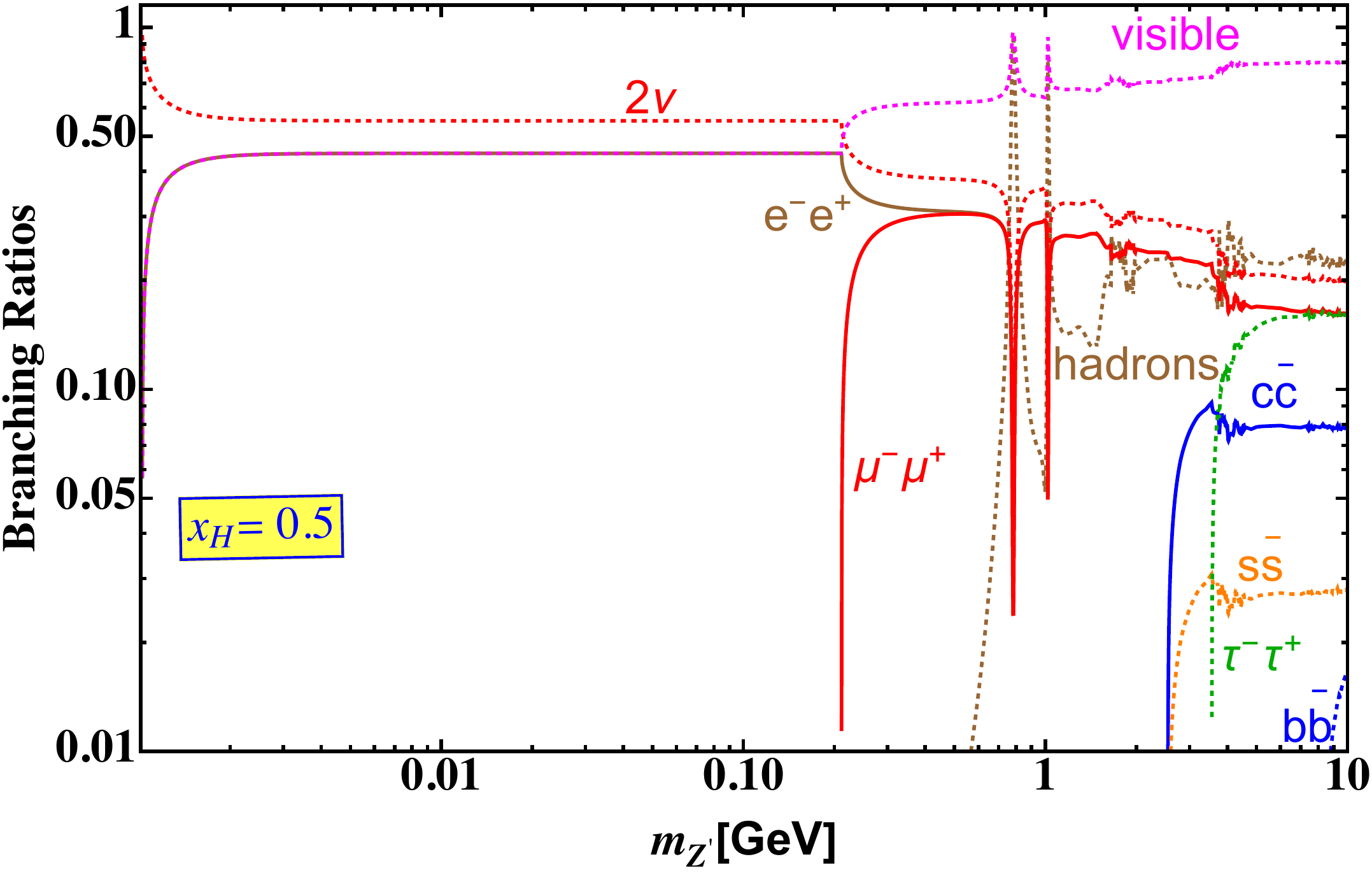}
\includegraphics[width=0.495\textwidth]{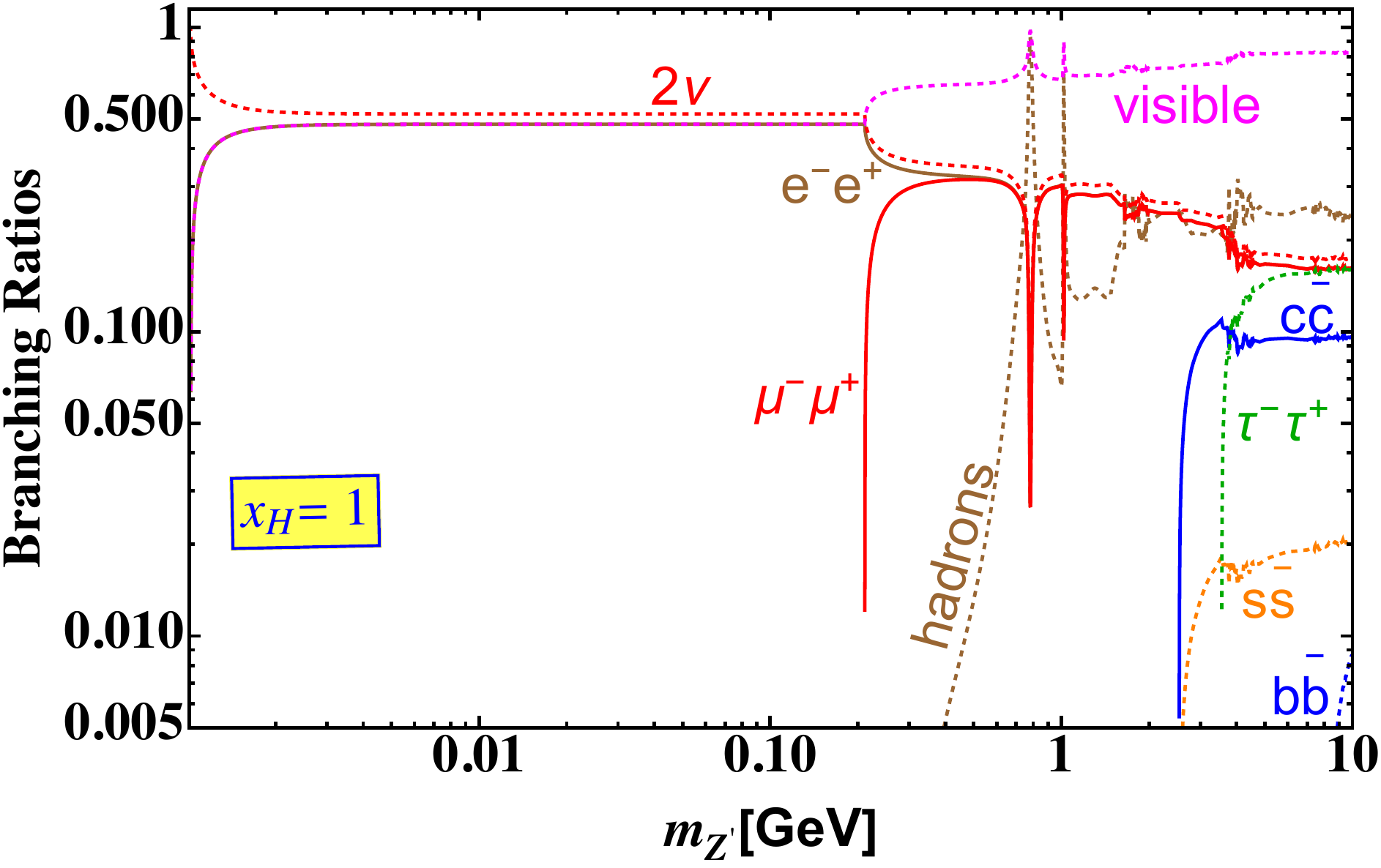}
\includegraphics[width=0.495\textwidth]{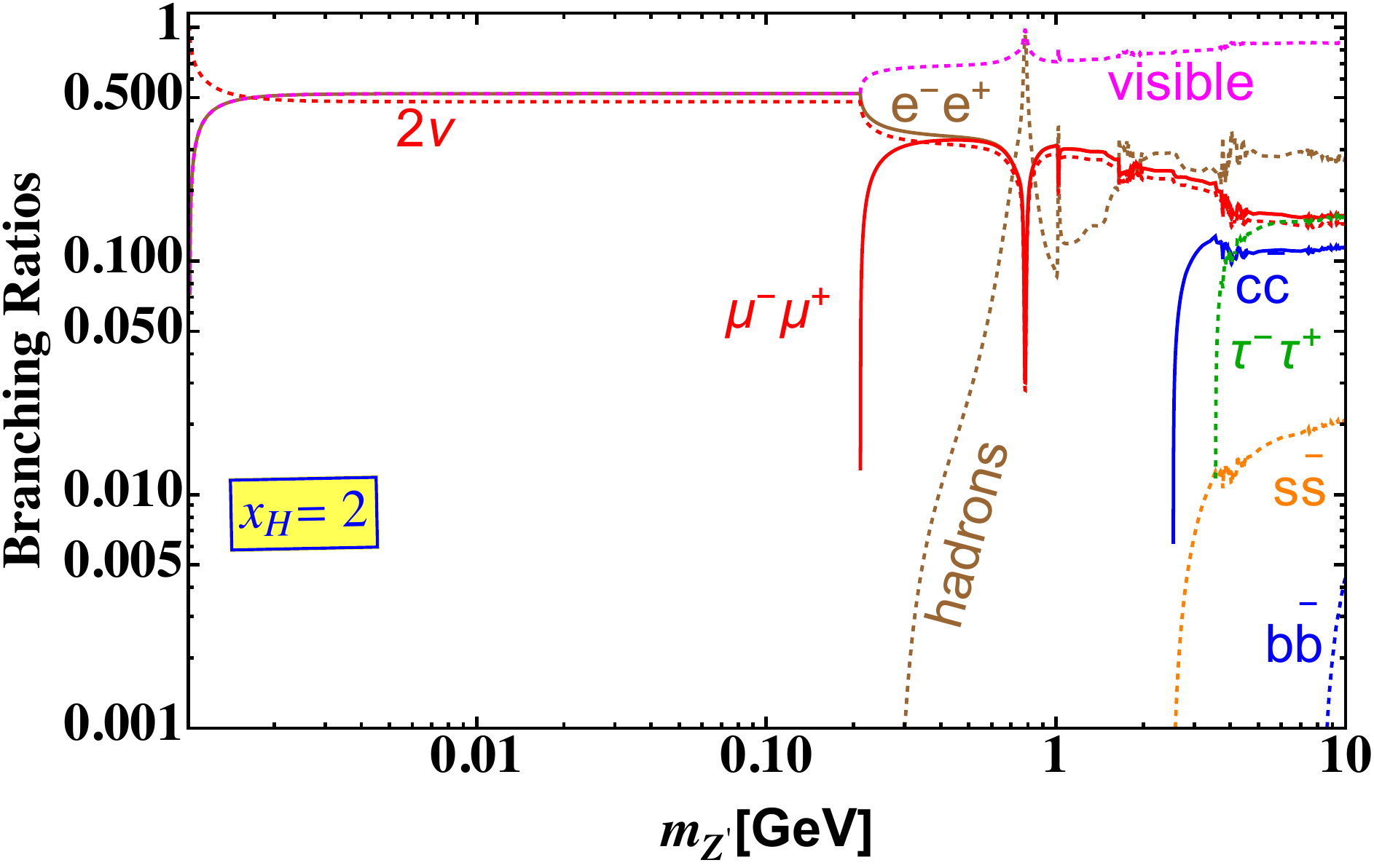}
\caption{Branching ratios of $Z^\prime$ into different modes for different $x_H^{}$ fixing $x_\Phi^{}=1$. For $x_H^{}=-2$, $Z^\prime$ decays only into visible modes, whereas in the case of the other choices of $x_H^{}$, $Z^\prime$ decays into both visible and invisible modes.}
\label{Zp1}
\end{figure}

Depending on the $Z^\prime$ mass in our analysis to probe $Z'$ in the context of FASER, FASER2, DUNE, and ILC beam dump experiments, we consider hadronic decay modes of $Z^\prime$. 
Because of the presence of U$(1)_X$ charges, we estimate the branching ratios of the $Z^\prime$ into different modes and show them in Figs.~\ref{Zp1} for different $x_H^{}$.
The partial decay width into the hadronic modes is complicated and calculated by the modified $R$ ratio for $m_{Z'}^{} \gtrsim 1.65$ GeV~\cite{Bauer:2018onh}.
On the other hand, for the lighter mass, vector-meson dominance (VMD) is valid, and we use the \texttt{DARKCAST} code from Ref.~\cite{Ilten:2018crw}. We find that for $x_H^{}=-2$ there is no interaction between the left-handed fermions and $Z^\prime$, which is a U$(1)_R$ scenario. 
As a result, the $Z^\prime$ can not decay into invisible mode. 
Therefore, in this case, $Z^\prime$ decays into visible mode only. 
For the other charges under consideration, we find that the left- and right-handed changed fermions have variations in interacting with the $Z^\prime$, which has been manifested in the branching ratios showing its chiral nature. 
It is important to mention that $x_H^{}=0$ is the U(1)$_{B-L}$ case which is a vector-like scenario.  

\section{Calculation of the constraints on the gauge couplings}
\label{sec:calc}

In beam dump experiments and FASER(2), $Z'$ gauge boson is produced through the bremsstrahlung process and rare decays of $\pi^0$ and $\eta$ mesons.
Moreover, for the cases of electron and positron beam dump experiments, the pair-annihilation process also contributes to the production of $Z'$ gauge bosons.
In this section, we explain the calculation of the number of signal events at proton, electron, and positron beam dump experiments and FASER(2).
The number of signals is roughly calculated by the number of produced $Z'$ gauge bosons and acceptance. 

Even if a lot of long-lived particles with short lifetimes or large angles with respect to a beam axis are produced at a beam dump, they decay in front of the detector or swerve from it.
Therefore, it is important to consider the acceptance of detectors for long-lived particle searches.
The acceptance of the detector is estimated by~\footnote{
For precise estimation, we need full Monte Carlo simulations to evaluate the acceptance of the detector, but it is left to future works.
}
\begin{align}
\label{eq:acceptance}
    {\rm Acc}(Z') =
    \mathcal{P}_{\rm dec}(m_{Z'}^{}, g_X^{}, \vec{p}_{Z'}^{}) \cdot \Theta_{\rm geom} \left( \frac{r_{\rm det}}{\ell_{\rm dump} + \ell_{\rm sh} + \ell_{\rm dec}} - \frac{p_{Z',{\rm lab}}^{\perp}}{p_{Z',{\rm lab}}^{z}} \right)~,
\end{align}
with $\ell_{\rm dump}$, $\ell_{\rm sh}$, and $\ell_{\rm dec}$ being the length of the beam dump, shield, and decay volume, respectively, $r_{\rm det}$ the radius of the detector, $p_{Z',{\rm lab}}^{\perp}$ the transverse momentum of $Z'$ gauge boson in the lab frame, and $p_{Z',{\rm lab}}^{z}$ that in the direction of the beam.
In Eq.~\eqref{eq:acceptance}, $\Theta_{\rm geom}(x)$ stands for the step function.
The probability that $Z'$ decays in the decay volume is calculated by
\begin{align}
\label{eq:decay-prob}
    \mathcal{P}_{\rm dec}(m_{Z'}^{}, g_X^{}, \vec{p}_{Z'}^{}) =
    \exp \left( - \frac{\ell_{\rm dump} + \ell_{\rm sh}}{\bar{d}_{Z',{\rm lab}}} \right) \left[ 1 - \exp \left( - \frac{\ell_{\rm dec}}{\bar{d}_{Z',{\rm lab}}^{}} \right) \right]~,
\end{align}
where $\bar{d}_{Z',{\rm lab}}^{}$ denotes the decay length of $Z'$ in the lab frame and is evaluated by
\begin{align}
\label{eq:decay-length}
    \bar{d}_{Z',{\rm lab}}^{} =
    \frac{|\vec{p}_{Z'}^{}|}{m_{Z'}^{}} \frac{1}{\Gamma_{\rm total}}~,
\end{align} 
with $\vec{p}_{Z'}^{}$ the momentum of $Z'$ gauge boson.

On the other hand, the ways to calculate the number of produced $Z'$ depend on the production processes of $Z'$. 
Hereafter, we review the calculations for the number of produced $Z'$ in each type of beam dump.

\subsection{Proton beam dump experiments and FASER}
\label{subsec:proton-beamdump}

In proton beam dump experiments and FASER, $Z'$ gauge bosons are mainly produced by the bremsstrahlung process and rare decays of $\pi^0$ and $\eta$ meson.
For the bremsstrahlung process, the production cross section is calculated by~\cite{Kim:1973he,Feng:2017uoz,Bauer:2018onh}~\footnote{
This subsection follows Ref.~\cite{Feng:2017uoz}, and the detailed discussion is shown in this reference.
}
\begin{align}
\label{eq:xsec-pbrems}
    \sigma(p p \to p Z' X) =
    \int \dd |\vec{p}_{Z'}^{}|^2 \int \dd \cos\theta_{Z'}^{} \frac{|\vec{p}_{Z'}^{}|}{|\vec{p}_{p_i}^{}|} w(|\vec{p}_{Z'}^{}|^2, \cos\theta_{Z'}^{}) \sigma_{pp}(s')~,
\end{align}
where $\theta_{Z'}^{}$ stands for the angle of $Z'$ respect to the beam axis, $\vec{p}_{p_i}^{}$ does for the momentum of initial proton in the beam, and $s' = 2 m_p (E_p - E_{Z'}^{})$ with $m_p$ being the proton mass and $E_p~(E_{Z'}^{})$ the energy of initial proton ($Z'$ gauge boson).
In the calculation of the cross section for bremsstrahlung production, we use the Fermi-Weizs\"{a}cker-Williams approximation~\cite{Fermi:1924tc,Williams:1934ad,vonWeizsacker:1934nji}, and $2 \to 3$ process is approximated by the $2 \to 2$ one,
\begin{align}
    p_i(p_{p_i}) + \gamma^*(q) \to p_f(p_{p_f}) + Z'(p_{Z'}^{})~,
\end{align}
where $\gamma^*$ stands for the virtual photon with momentum $q$, and this approximation is valid when the energies of relating particles are much higher than their masses and transverse momentum of $Z'$ as $E_{Z'}^{}, E_{p_i}, E_{p_f} \gg m_{Z'}^{}, m_p, p_{Z',{\rm lab}}^\perp$~\cite{Blumlein:2013cua}.
Cross section of $pp$ inelastic scattering is denoted by $\sigma_{pp}(s)$, and $w(|\vec{p}_{Z'}^{}|^2, \cos\theta_{Z'}^{})$ is the splitting function~\cite{Kim:1973he,Tsai:1973py,Blumlein:2013cua} given by
\begin{align}
\label{eq:split}
    w(|\vec{p}_{Z'}^{}|^2, \cos\theta_{Z'}^{}) =
    &\frac{g_X^2}{16 \pi^2 H} \left[ (x_{p,L}^{\prime 2} + x_{p,R}^{\prime 2}) \frac{1 + (1-z)^2}{z} + (x_{p,L}^\prime - x_{p,R}^\prime)^2 z \frac{m_p^2}{m_{Z'}^2} \right. \nonumber \\
    &\qquad \qquad \left. + 2 z (1-z) \frac{ (x_{p,L}^{\prime 2} - 6 x_{p,L}^\prime x_{p,R}^\prime + x_{p,R}^{\prime 2}) m_p^2 - (x_{p,L}^{\prime 2} + x_{p,R}^{\prime 2}) m_{Z'}^2 }{H}  \right. \nonumber \\
    &\qquad \qquad + 2 z (1-z) \left\{ z^2 (x_{p,L}^{\prime 2} + x_{p,R}^{\prime 2}) m_p^4 + (1-z) (x_{p,L}^{\prime 2} + x_{p,R}^{\prime 2}) m_{Z'}^4  \right. \nonumber \\
    &\qquad \qquad \left. \left. + \left( (z^2 + z -1) (x_{p,L}^{\prime 2} + x_{p,R}^{\prime 2}) + 6 (1-z) x_{p,L}^\prime x_{p,R}^\prime \right) m_p^2 m_{Z'}^2 \right\} \frac{1}{H^2} \right]~,
\end{align}
where
\begin{align}
    z 
    &= \frac{p_{Z',{\rm lab}}^z}{|\vec{p}_{p_i}|}
    = \frac{|\vec{p}_{Z'}^{}|}{|\vec{p}_{p_i}|} \cos\theta_{Z'}~, \\
    H
    &= (p_{Z'}^\perp)^2 + (1-z) m_{Z'}^2 + z^2 m_p^2
    = |\vec{p}_{Z'}^{}|^2 (1 - \cos^2\theta_{Z'}) + (1-z) m_{Z'}^2 + z^2 m_p^2~.
\end{align}
The U(1)$_X$ charge of the proton are obtained by $x_{p,L}^\prime = 3 x_q^\prime$ and $x_{p,R}^\prime = 2 x_u^\prime + x_d^\prime$.
According to Refs.~\cite{Kim:1973he,Tsai:1973py}, the integrand in Eq.~\eqref{eq:xsec-pbrems} more dominantly contributes to the cross section when the virtuality of the virtual photon $\gamma^*$ is smaller. 
This virtuality is evaluated by
\begin{align}
\label{eq:qmin}
    |q_{\rm min}^2| \approx
    \frac{1}{4 E_{p_i} z^2 (1-z)^2} \left\{ (p_{Z',{\rm lab}}^\perp)^2 + (1-z) m_{Z'}^2 + z^2 m_p^2 \right\}^2~,
\end{align}
and we apply the momentum cut as $|q_{\rm min}^2| < \Lambda_{\rm QCD}^{}$ with $\Lambda_{\rm QCD}^{} \simeq 250$ MeV being the QCD scale, following Ref.~\cite{Feng:2017uoz}.
It should be emphasized that the second term in the square bracket of Eq.~\eqref{eq:split} is enhanced when $x_{p,L}^\prime \neq x_{p,R}^\prime$ and $Z'$ is much lighter than the proton, and this is a characteristic behavior of the chiral $Z'$ gauge boson.

For the $Z'$ gauge boson with a mass as heavy as vector mesons, the electromagnetic form factor of nucleons cannot be ignored.
In the VMD model mentioned in the last of Sec.~\ref{sec:model}, this form factor, which incorporates both the Breit-Wigner components of the $\rho$- and $\omega$-like mesons, is calculated by~\cite{Faessler:2009tn,deNiverville:2016rqh}
\begin{align}
\label{eq:form-factor}
    F_1(p_{Z'}^2) =
    \sum_{V = \rho, \rho', \rho'', \omega, \omega', \omega''} \frac{f_V^{} m_V^2}{m_V^2 - |\vec{p}_{Z'}|^2 - i m_V^{} \Gamma_V^{}}~,
\end{align}
where $m_V^{} (\Gamma_V^{})$ stands for the mass (decay width) of the meson $V$.
In Table~\ref{tab:form-factor}, the values used in the calculation of the form factor are shown.
\begin{table}[t]
\begin{center}
\begin{tabular}{|c|cccccc|}
\hline
Meson $V$ & $\rho$ & $\rho'$ & $\rho''$ & $\omega$ & $\omega'$ & $\omega''$ \\
\hline
$f_V^{}$ & 0.616 & 0.223 & -0.339 & 1.011 & -0.881 & 0.369 \\
$m_V^{}$ [GeV] & 0.770 & 1.250 & 1.450 & 0.770 & 1.250 & 1.450 \\
$\Gamma_V^{}$ [GeV] & 0.150 & 0.300 & 0.500 & 0.0085 & 0.300 & 0.500 \\
\hline
\end{tabular}
\end{center}
\caption{
Values used in the calculation of the form factor in Eq.~\eqref{eq:form-factor}.
These values have been taken from Refs~\cite{Feng:2017uoz,Faessler:2009tn}.
}
\label{tab:form-factor}
\end{table}
The differential production rate of the $Z'$ gauge boson is calculated by
\begin{align}
\label{eq:prod-rate_p-brem}
    \frac{\dd N^{\rm brem}}{\dd|\vec{p}_{Z'}^{}|^2 \dd \cos\theta_{Z'}^{}} =
    \frac{|\vec{p}_{Z'}^{}|}{|\vec{p}_{p_i}|} w(|\vec{p}_{Z'}^{}|^2, \cos\theta_{Z'}^{}) \frac{\sigma_{pp}(s')}{\sigma_{pp}(s)}~,
\end{align}
with $s = 2 m_p E_{p_i}$.

On the other hand, for production by meson rare decays, the differential production rate is calculated by
\begin{align}
\label{eq:prod-rate_p-meson}  
    \frac{\dd N^M}{\dd|\vec{p}_M^{}|^2 \dd \cos\theta_M^{}} =
    \frac{d\sigma(p p \to M X)}{\dd|\vec{p}_M^{}|^2 \dd \cos\theta_M^{}} \cdot {\rm BR}(M \to Z' \gamma)~,
\end{align}
where $\vec{p}_M^{}$ and $\theta_M^{}$ are the momentum and angle of meson respect to the beam axis, respectively.
The branching fractions of the rare meson decays are given by
\begin{align}
    {\rm BR}(\pi^0 \to Z' \gamma) &=
    2 \left( 1 - \frac{m_Z^2}{m_{\pi^0}^2} \right)^3 {\rm BR}(\pi^0 \to \gamma \gamma) \cdot R_{\pi^0}(m_{Z'})~, \\
    {\rm BR}(\eta \to Z' \gamma) &=
    2 \left( 1 - \frac{m_Z^2}{m_{\eta}^2} \right)^3 {\rm BR}(\eta \to \gamma \gamma) \cdot R_{\eta}(m_{Z'})~,
\end{align}
where ${\rm BR}(\pi^0 \to \gamma \gamma) \simeq 0.99$ and ${\rm BR}(\eta \to \gamma \gamma) \simeq 0.39$.
The function, $R_{\pi^0 (\eta)}(m_{Z'}^{})$, is the ratio of decay widths between $\pi^0 (\eta) \to Z' \gamma$ and $\pi^0 (\eta) \to A' \gamma$ with $A'$ being the dark photon and is calculated by \texttt{darkcast}~\cite{Ilten:2018crw}.

From Eqs.~\eqref{eq:prod-rate_p-brem} and \eqref{eq:prod-rate_p-meson}, the total expected numbers of events in proton beam dump experiments are given by
\begin{align}
\label{eq:num_p-brem}
    N_{\rm event}^{\rm p\mathchar`-brem} &=
    N_p\, |F_1(m_{Z'}^2)|^2 \int \dd |\vec{p}_{Z'}^{}|^2 \int \dd \cos\theta_{Z'}^{}\, \frac{\dd N}{\dd |\vec{p}_{Z'}^{}|^2 \dd \cos\theta_{Z'}^{}}\, \Theta(\Lambda_{\rm QCD}^{} - q^2) \cdot {\rm Acc}(Z')~, \\
\label{eq:num_p-meson}
    N_{\rm event}^{\rm p\mathchar`-meson} &=
    N_p\, \sum_{M = \pi^0, \eta} \int \dd |\vec{p}_M^{}|^2 \int \dd \cos\theta_M^{} \int \dd |\vec{p}_{Z'}^{}|^2 \int \dd \cos\theta_{Z'}^{}\, \frac{\dd N^M}{\dd|\vec{p}_M^{}|^2 \dd \cos\theta_M^{}} \cdot {\rm Acc}(Z')~,
\end{align}
where $N_p$ stands for the number of protons on target.

For FASER(2) case, the total expected number of events can be calculated in the almost same way as in the case of proton beam dumps.
It should be noted that the only difference is that both protons have high energies in the lab frame.
Therefore, for the calculation by using Eq.~\eqref{eq:num_p-brem}, the Lorentz transformation needs to obtain the momentum and energy of one proton in the rest frame of the other one. 

In this paper, we take account of $\nu$-Cal~\cite{Blumlein:2011mv,Blumlein:2013cua}, LSND~\cite{LSND:1997vqj}, PS191~\cite{Bernardi:1985ny}, NOMAD~\cite{NOMAD:2001eyx}, and CHARM~\cite{CHARM:1985anb} as past proton beam dump experiments and DUNE~\cite{DUNE:2016hlj,DUNE:2015lol,DUNE:2021tad} as future one.
Moreover, we also calculate sensitivity regions by FASER(2), which is an experiment for the search of long-lived particles produced by proton-proton collision at the ATLAS interaction point.
For $\nu$-Cal, DUNE, and FASER(2) cases, we obtain bound and sensitivity regions for the chiral $Z'$ gauge boson by Eqs.~\eqref{eq:num_p-brem} and \eqref{eq:num_p-meson}.
The specification of the experiments is shown in Table~\ref{tab:experiment}.
\begin{table}[t]
\begin{center}
\begin{tabular}{|c|cccccccc|}
\hline
Experiment & Beam type & $E_{\rm beam}$ [GeV] & $N_{e^\pm}, N_p$ & Target [${}_Z^AX$] & $(\ell_{\rm dump} + \ell_{\rm sh})$ [m] & $\ell_{\rm dec}$ [m] & $r_{\rm det}$ [m] & $N_{95\%}$ \\ \hline
E137 & $e^-$ & 20 & $1.87 \times 10^{20}$ & ${}_{13}^{26.98}{\rm Al}$ & 179 & 204 & 1.5 & 3 \\
E141 & $e^-$ & 9 & $2 \times 10^{15}$ & ${}_{74}^{183.84}{\rm W}$ & 0.12 & 0.35 & 0.075 & 3419 \\
ILC beam dump & $e^+$ & 125 & $4 \times 10^{22}$ & ${\rm H_2O}$ & 81 & 50 & 2 & 3 \\
$\nu$-Cal & $p$ & 68.6 & $1.71 \times 10^{18}$ & ${}_{26}^{55.85}{\rm Fe}$ & 64 & 23 & 1.3 & 4.5 \\
DUNE & $p$ & 120 & $1.47 \times 10^{22}$ & ${}_{6}^{12.01}{\rm C}$ & 574 & 5 & 2.5 & 3 \\
 & & & & & & & & \\ \cline{3-4}
 & & $\sqrt{s}$ [TeV] & $\mathcal{L}$ [fb$^{-1}$] & & & & & \\ \cline{3-4}
FASER & $p$ & 13 & 150 & $p$ & 478.5 & 1.5 & 0.1 & 3 \\
FASER2 & $p$ & 14 & 3000 & $p$ & 475 & 5 & 1 & 3 \\
\hline
\end{tabular}
\end{center}
\caption{
Specification of the beam dump experiments and FASER(2).
Symbols for chemical elements with the atomic (mass) number, $Z (A)$, are denoted by $X$.
$s$ and $\mathcal{L}$ stand for the center of mass energy of proton-proton collision at the ATLAS interacting point and integrated luminosity, respectively.
$95\%$ exclusion sensitivity corresponds to $N_{\rm event} \geq N_{95\%}$.
}
\label{tab:experiment}
\end{table}

For LSND, PS191, NOMAD, and CHARM cases, we obtain bound curves by rescaling the bounds of U(1)$_{B-L}$ case~\cite{Bauer:2018onh}. 
Then the upper bound on $\{m_{Z'}^{}, g_X^{} \}$ plane is approximately scaled applying~\cite{Ilten:2018crw,Chakraborty:2021apc} 
\begin{equation}
\label{eq:upper}
\tau_{Z'}^{}(g_{B-L}^{\rm max}) \sim \tau_{Z'}^{}(g_X^{\rm max}, x_H^{}, x_\Phi^{})~,
\end{equation} 
where $\tau_{Z'}^{}$ is the lifetime of the $Z'$, and $g_{B-L}^{}$ denotes the gauge coupling in the U(1)$_{B-L}$ case.
The lower bound is scaled by 
\begin{equation}
\label{eq:lowerP}
g_X^{\rm low} \sim g_{B-L}^{\rm low} \sqrt{ \frac{{\rm BR}(M \to Z'_{B-L} \gamma)~ {\rm BR}(Z'_{B-L} \to e^+ e^-) \tilde \tau_{Z'}^{} }{{\rm BR}(M \to Z' \gamma)~ {\rm BR}(Z' \to e^+ e^-) \tilde \tau_{Z'_{B-L} }} }~,
\end{equation}
where $\tilde \tau$ is lifetime with gauge coupling being unity and
$Z'$ is produced via meson decay with 
$M = \pi^0$ for LSND, PS191, and MONAD, and $M = \eta$ for CHARM. 
Here, the branching ratio of meson decay is estimated using the 
method given in Ref.~\cite{Ilten:2018crw}.

\subsection{Electron and positron beam dump experiments}
\label{subsec:electron-beamdump}

In electron and positron beam dump experiments, $Z'$ gauge bosons are mainly produced by not only the bremsstrahlung process and rare decay of $\pi^0$ and $\eta$ mesons, but also pair annihilation. 
One difference from proton beam dump experiments is that electrons and positrons in the beam produce electromagnetic showers in the beam dump, and thus the initial particles have various energies and angles with respect to the beam axis.

For the bremsstrahlung process, the production cross section is calculated as~\cite{Bjorken:2009mm,Liu:2016mqv,Liu:2017htz,Asai:2021ehn}
\begin{align}
\label{eq:xsec-ebrems}
    \frac{\dd \sigma(e^\pm {\rm N} \to e^\pm {\rm N} Z')}{\dd x \dd \theta_{Z'}^{}} =
    \frac{g_X^2 \alpha_{\rm em}^2}{2 \pi} x (1-x) E_i^2 \beta_{Z'}^{} \frac{\mathcal{A}^{Z'}|_{t=t_{\rm min}}}{\tilde{u}^2} \chi~,
\end{align}
where $\alpha_{\rm em}$ stands for the fine structure constant, $x = E_{Z'}^{}/E_i$, $\beta_{Z'}^{} = \sqrt{1 - m_{Z'}^2/E_i^2}$, and
\begin{align}
    \tilde{u} =
    - x E_i^2 \theta_{Z'}^2 - m_{Z'}^2 \frac{1-x}{x} - m_e^2 x~.
\end{align}
The effective photon flux, $\chi$, is calculated as~\cite{Bjorken:2009mm}
\begin{align}
\label{eq:eff-photon-flux}
    \chi = \int_{t_{\rm min}}^{t_{\rm max}} \dd t \frac{t - t_{\rm min}}{t^2} G_2(t)~,
\end{align}
where $t_{\rm min} = \left\{ m_{Z'}^2 / (2 E_i) \right\}^2$ and $t_{\rm max} = m_{Z'}^2$.
The electron form factor, $G_2(t)$, is given by the elastic and inelastic component as follows:
\begin{align}
\label{eq:G2}
    G_2(t) = G_{2,{\rm el}}(t) + G_{2,{\rm inel}}(t)~,
\end{align}
and these two components are given by
\begin{align}
\label{eq:G2-elastic}
    G_{2,{\rm el}} &=
    \left( \frac{a^2 t}{1 + a^2 t} \right)^2 \left( \frac{1}{1 + t/d} \right)^2 Z^2~, \\
\label{eq:G2-inelastic}
    G_{2,{\rm inel}} &=
    \left( \frac{a^{'2} t}{1 + a^{'2} t} \right)^2 \left( \frac{1 + \frac{t}{4 m_p^2} (\mu_p^2 - 1)}{ \left(1 + \frac{t}{0.71\,{\rm GeV}^2} \right)^4} \right)^2 Z~,
\end{align}
where $a = 111 Z^{-1/3} / m_e$, $d = 0.164\,{\rm GeV}^2 A^{-2/3}$, $a' = 773 Z^{-2/3} / m_e$, and $\mu_p = 2.79$ with $Z$ and $A$ being the atomic and mass number of the beam dump material, respectively.
The amplitude in Eq.~\eqref{eq:xsec-ebrems}, $\mathcal{A}^{Z'}$, are obtained as follows~:
\begin{align}
    \mathcal{A}^{Z'} |_{t = t_{\rm min}} &=
    (x_\ell^{\prime 2} + x_e^{\prime 2}) \frac{2-2x+x^2}{1-x} + 2 \left( (x_\ell^{\prime 2} + x_e^{\prime 2}) m_{Z'}^2 + (x_\ell^{\prime 2} - 6 x_\ell^\prime x_e^\prime + x_e^{\prime 2}) m_e^2 \right) \frac{x}{\tilde{u}} \nonumber \\
    &\quad + 2 \left\{ (x_\ell^{\prime 2} + x_e^{\prime 2}) (1-x) m_{Z'}^4 - \left( x_\ell^{\prime 2} (1-x-x^2) - 6 x_\ell^\prime x_e^\prime (1-x) + x_e^{\prime 2} (1-x-x^2) \right) m_{Z'}^2 m_e^2 \right. \nonumber \\
    &\qquad \qquad \left. - (x_\ell^{\prime 2} - 6 x_\ell^\prime x_e^\prime + x_e^{\prime 2}) x^2 m_e^4 \right\} \frac{1}{\tilde{u}^2}~.
\end{align}
The number and energy distribution of the electrons and positrons in the beam dump are evaluated by the track length, $l_i\, (i = e^+, e^-)$, discussed in Appendix~\ref{app:track-length}.

For the pair annihilation process, the production cross section is calculated as~\cite{Marsicano:2018krp,Asai:2021ehn}
\begin{align}
\label{eq:xsec-eanni}
    \sigma(e^+e^- \to Z') &=
    \frac{1}{2} m_{Z'}^2 g_X^2 (x_\ell^{\prime 2} + x_e^{\prime 2}) \frac{\Gamma_{Z'}^{}/4}{(\sqrt{s} - m_{Z'})^2 + \Gamma_{Z'}^2/4} \nonumber \\
    &\simeq \frac{\pi g_X^2}{4 m_e} (x_\ell^{\prime 2} + x_e^{\prime 2})\, \delta\left( E_i - \frac{m_{Z'}^2}{2 m_e} + m_e \right)~,
\end{align}
where $s$ stands for the center-of-mass energy squared, and $\delta(x)$ is the Dirac delta function. 
In Eq.~\eqref{eq:xsec-eanni}, we use the narrow-width approximation.

Lastly, for production by meson rare decay, the differential production rate is calculated in the same way as the proton beam dump case and given by
\begin{align}
\label{eq:prod-rate_e-meson}
    \frac{\dd N^M}{\dd|\vec{p}_M^{}|^2 \dd \cos\theta_M^{}} =
    \frac{d\sigma(e^\pm p \to M X)}{\dd|\vec{p}_M^{}|^2 \dd \cos\theta_M^{}} \cdot {\rm BR}(M \to Z' \gamma)~.
\end{align}

The total expected numbers of events in electron and positron beam dump experiments are given by
\begin{align}
\label{eq:num_e-dump_brems}
    N_{\rm event}^{\rm brem} &=
    N_{e^\pm} \sum_{i = e^-, e^+}\int \dd E_i \frac{\dd l_i}{\dd E_i} \cdot n_{\rm N}^{} \int \dd E_X^{} \int \dd \theta_X^{} \frac{\dd \sigma(e^\pm {\rm N} \to e^\pm {\rm N} Z')}{\dd x \dd \theta_{Z'}^{}} \cdot {\rm Acc}(Z')~, \\
\label{eq:num_e-dump_anni}
    N_{\rm event}^{\rm anni} &=
    N_{e^\pm} \int \dd E_{e^+} \frac{\dd l_{e^+}}{\dd E_{e^+}} \cdot n_{e^-} \cdot \sigma(e^+e^- \to Z') \cdot {\rm Acc}(Z')~, \\
\label{eq:num_e-meson}
    N_{\rm event}^{\rm p\mathchar`-meson} &=
    N_{e^\pm}\, \sum_{M = \pi^0, \eta} \int \dd |\vec{p}_M^{}|^2 \int \dd \cos\theta_M^{} \int \dd |\vec{p}_{Z'}^{}|^2 \int \dd \cos\theta_{Z'}^{}\, \frac{\dd N^M}{\dd|\vec{p}_M^{}|^2 \dd \cos\theta_M^{}} \cdot {\rm Acc}(Z')~,
\end{align}
where $N_{e^\pm}$ is the number of incident electrons and positrons into the beam dump, and $n_{\rm N}^{} (n_{e^-})$ is the number density of the nucleus (electron) for the target material.

In this paper, we take account of E137~\cite{Bjorken:1988as} E141~\cite{Riordan:1987aw}, Orsay~\cite{Davier:1989wz}, and KEK~\cite{Konaka:1986cb} as past electron beam dump experiments and ILC beam dump experiment~\cite{Sakaki:2020mqb,Asai:2021xtg,Asai:2021ehn,Moroi:2022qwz} as future one.
For E137, E141, and ILC beam dump experiment cases, we obtain bound and sensitivity regions for the chiral $Z'$ gauge boson by Eqs.~\eqref{eq:num_e-dump_brems}, \eqref{eq:num_e-dump_anni}, and \eqref{eq:num_e-meson}.
The specification of the experiments are shown in Table~\ref{tab:experiment}.

For Orsay and KEK cases, we obtain bound curves by rescaling the bounds of U(1)$_{B-L}$ case as for the proton beam dump case. 
The constraint on the upper region on $\{m_{Z'}^{}, g_X^{} \}$ plane is approximately scaled by using Eq.~\eqref{eq:upper}, which is the same as the proton beam dump case.
On the other hand, the constraint on the lower region is obtained by
\begin{equation}
\label{eq:bremsstrahlung}
g_X^{\rm low} \sim g_{B-L}^{\rm low} \sqrt{ \frac{2 {\rm BR}(Z'_{B-L} \to e^+ e^-) \tilde \tau_{Z'}^{} }{(5x_H^{\prime 2}/4 + 3x_H^\prime x_\Phi^\prime + 2x_\Phi^{\prime 2}) {\rm BR}(Z' \to e^+ e^-) \tilde \tau_{Z'_{B-L}}^{}} }~,
\end{equation}
where $Z'$ is produced via the bremsstrahlung process.

\subsection{Supernova bound}
\label{subsec:sn}
The process of calculations of the constraints from SN1987A has been described in Ref.~\cite{Croon:2020lrf} for the B$-$L and flavored scenarios. 
In our case, we follow this method, however, due to the chiral nature, $Z^\prime$ will differently interact with the left- and right-handed fermions. 
Due to this fact, we can not directly use the results given in Ref.~\cite{Croon:2020lrf} and will calculate the necessary analytical expressions required for the chiral scenarios to estimate bounds on parameters of the general U$(1)_X^{}$ scenarios. 
Hence, we can systematically compare the bounds obtained from supernova with those obtained from FASER, FASER2, DUNE, and ILC beam dump experiments.

The luminosity of the light $Z'$ from the supernova is given by
\begin{align}
  L_{\rm LLP}^{} &= \int {\rm d}V \int \frac{{\rm d}^3 k}{(2\pi)^3}~ \omega \Gamma_{\rm prod}(\omega, r) \exp [ -\tau_{\rm abs} (\omega ,r) ]   \nonumber\\
  &=\int^{R_\nu}_0 (4 \pi r^2 {\rm d}r)\frac{1}{2 \pi^2} \int_{m_{Z'}}^{\infty} {\rm d}\omega~ \omega^2 \sqrt{\omega^2 - m^2_{Z'}} \Gamma_{\rm prod}(\omega, r) \exp [ -\tau_{\rm abs} (\omega ,r) ] ~,
\end{align}
where the parameter $r$ is the radial distance inside the neutrinosphere (with $R_\nu \sim 25$ km), $\omega$ is the energy of the $Z'$, and $m_{Z'}$ is the mass of the $Z'$. 
The rescattering and reabsorption effects of $Z'$ traveling through the star are described by the absorptive optical depth 
\begin{align}
\tau_{\rm abs}= \int {\rm d}\bar{r}~ \Gamma_{\rm abs}(\omega , \bar{r}) = \int_r^{R_{\rm far}} {\rm d}\bar{r}~ e^{\omega/T(\bar{r})} \Gamma_{\rm prod} (\omega, \bar{r})~,
\end{align}
where the second equality is due to principle of detailed balance~\cite{Weldon:1983jn}, $T(\bar{r})$ is the temperature of the star at radius $\bar{r}$~\cite{Bollig:2020xdr}, and $R_{\rm far}$ is taken to be 100 km. 

\begin{figure}[t]
\centering
\includegraphics[width=1\textwidth]{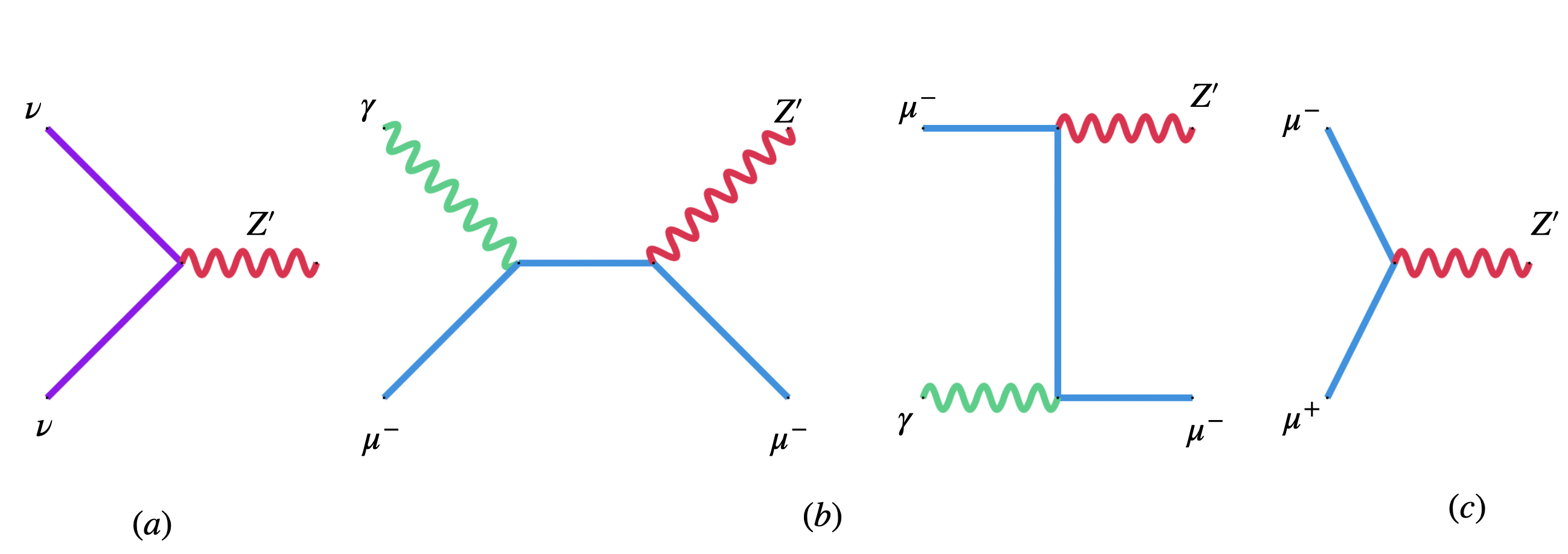}
\caption{The Feynman diagrams of (a) neutrino-pair coalescence, (b) semi-Compton scattering, and (c) muon-pair coalescence processes, respectively.}
\label{SN}
\end{figure}

\begin{figure}
\begin{center}
\includegraphics[width=0.49\textwidth]{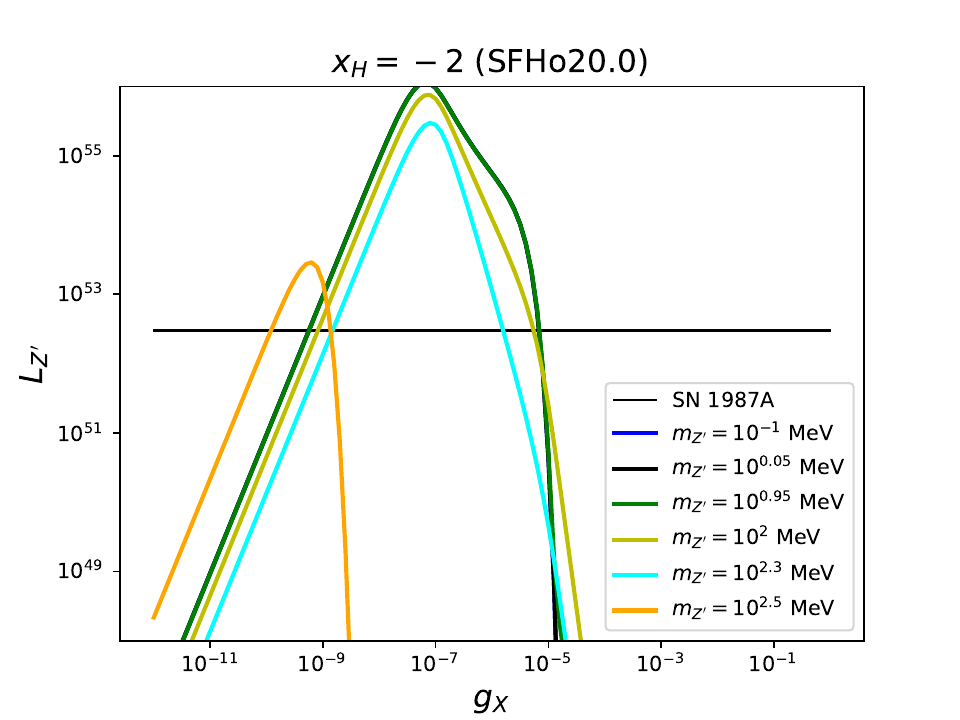}
\includegraphics[width=0.49\textwidth]{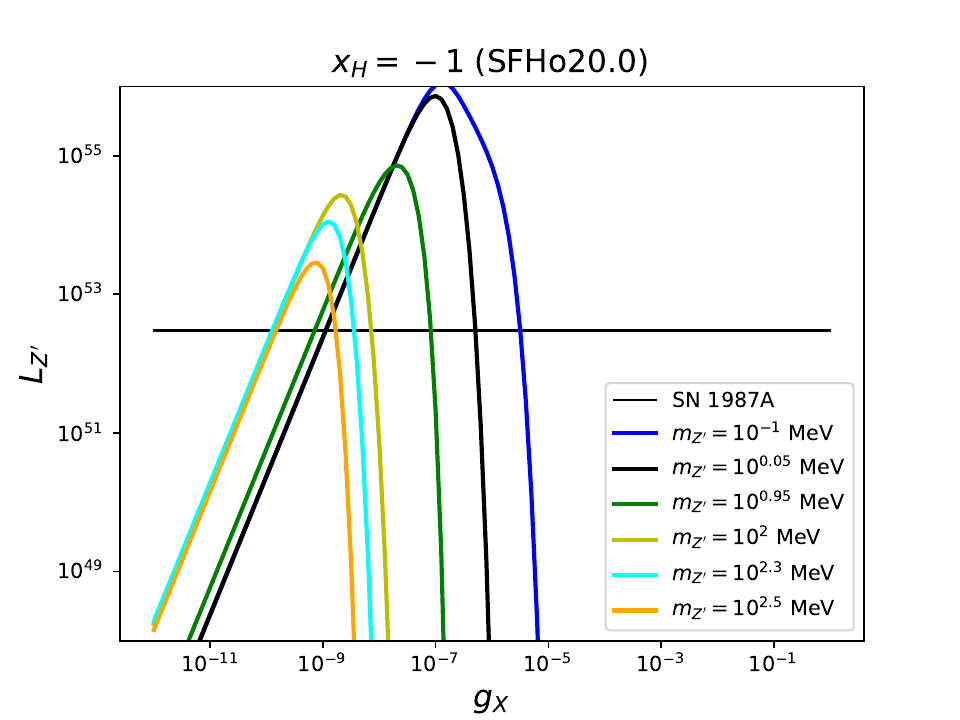}\\
\includegraphics[width=0.49\textwidth]{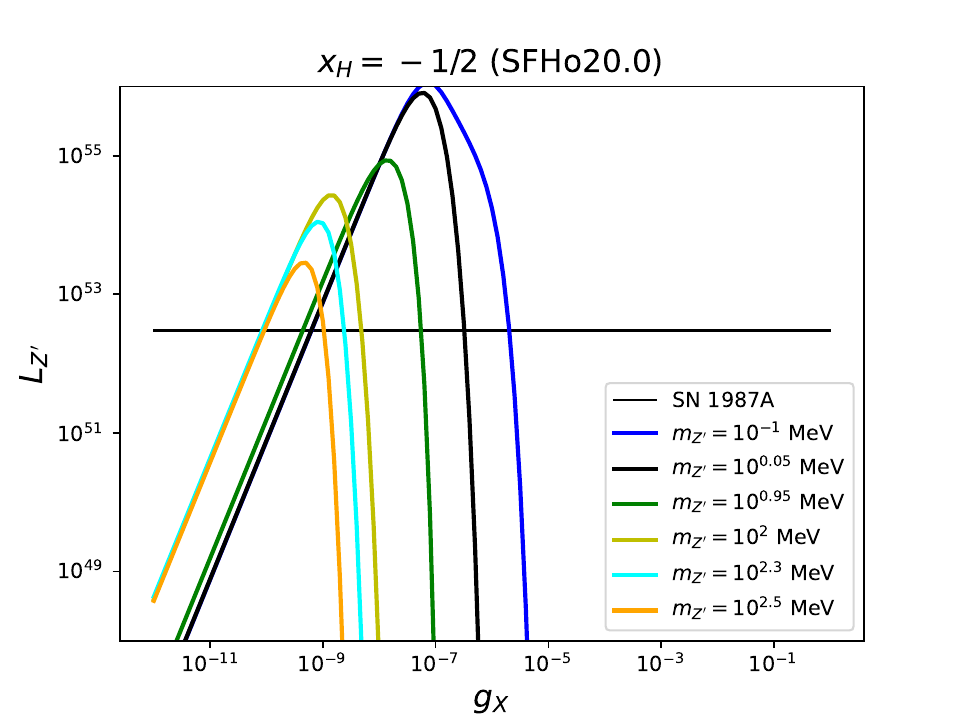}
\includegraphics[width=0.49\textwidth]{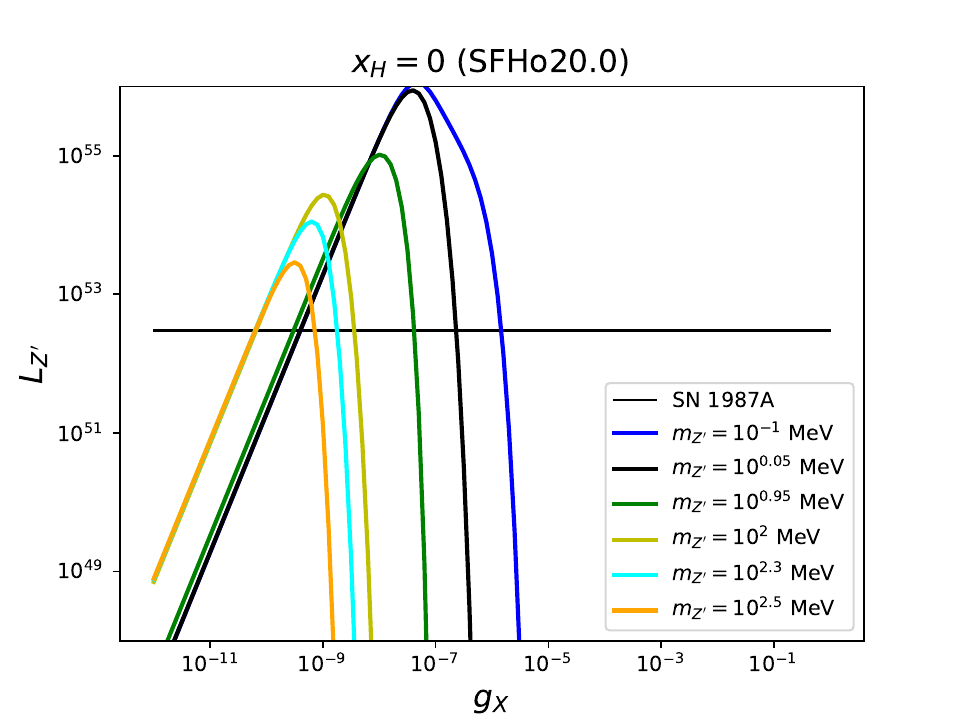}\\
\includegraphics[width=0.49\textwidth]{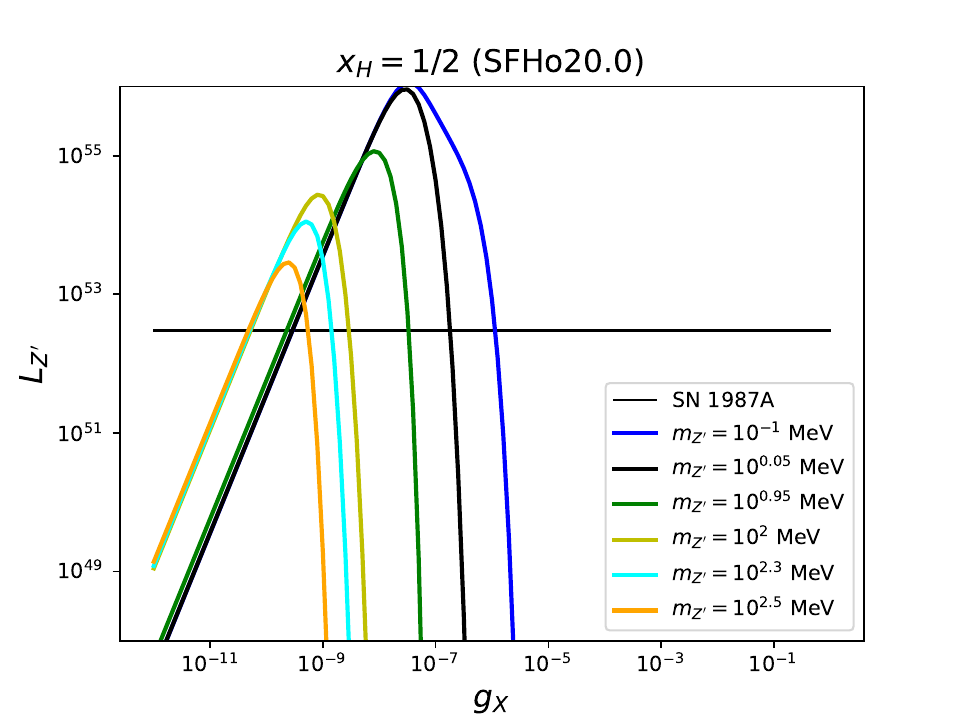}
\includegraphics[width=0.49\textwidth]{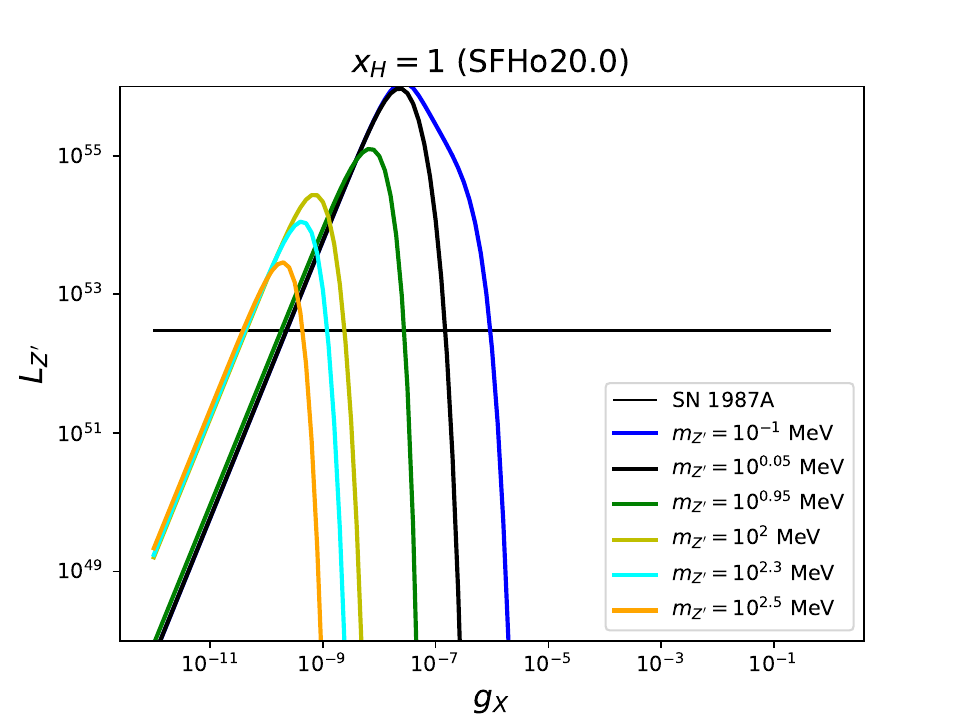}\\
\includegraphics[width=0.49\textwidth]{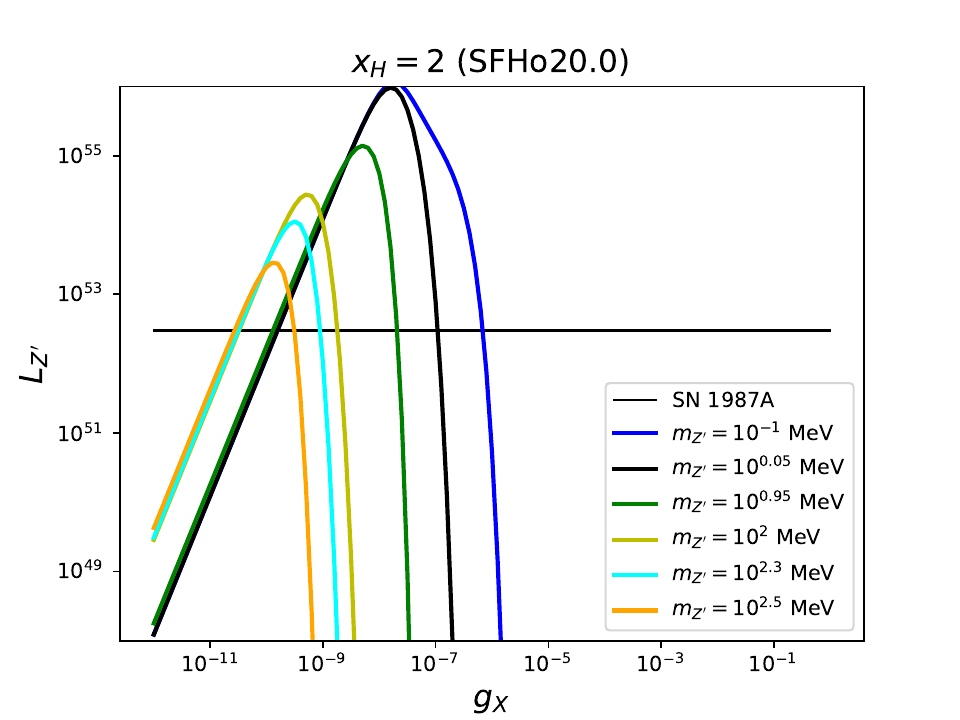}
\caption{Luminosity $(L_{Z^\prime}^{}$ in ergs/sec) due to the $Z^\prime$ production in a proto-neutron star as a function of the $g_X^{}$ under a general U(1)$_X$ scenario for SFHo20.0 profile. The SN1987A line in solid black (horizontal) represents luminosity carried away by the neutrinos. Different colors represent different $m_{Z^\prime}^{}$. The luminosity for SFHo18.8 profile will look the same, however, less strong.}
\label{Supernova1}
\end{center}
\end{figure}

In this work, we consider the $Z'$ productions from neutrino-pair coalescence ($\nu \nu \to Z'$), semi-Compton scattering ($\gamma \mu \to Z' \mu$), and muon-pair coalescence ($\mu \mu \to Z'$) as follows~:
\begin{align}
\Gamma_{\rm prod}(\omega, r) = \Gamma_{\nu \nu \to Z'} +\Gamma_{\gamma \mu \to Z' \mu}  + \Gamma_{\mu \mu \to Z'} ~,
\end{align}
and the relevant processes are shown in Fig.~\ref{SN}. 
The rate of the neutrino pair coalescence is given by
\begin{align}
\Gamma_{\nu \nu \to Z'} (\omega, r)  &=  \frac{g^2_{Z'\nu\nu} m^2_{Z'}}{3 \pi \omega} \times \frac{1}{e^{\omega/T} -1}~,
\end{align}
considering $m_{Z^\prime}^{} \gg 2 m_\nu$. The rate of the semi-Compton scattering is calculated by~\cite{Caputo:2021rux}
\begin{align}
\Gamma_{\gamma \mu \to Z' \mu} (\omega, r)  &= \sigma_T \frac{n_{\mu} (r) F_{\rm deg}(r)}{e^{\omega/T} -1} \sqrt{1- \frac{m^2_{Z'}}{\omega^2}} ~,
\end{align}
where 
\begin{align}
\sigma_T= \frac{4 g^2_{Z' \mu\mu} \alpha_{\rm em}}{3 m^2_\mu} \times \frac{3}{4} \left[ \frac{16}{(\hat{s}-1)^2} +\frac{\hat{s}+1}{\hat{s}^2}+ \frac{2(\hat{s}^2-6\hat{s}-3)}{(\hat{s}-1)^3} \ln \hat{s} \right]~,
\end{align}
with $\hat{s} \equiv 1+ 2\omega /m_\mu$. In the region of light $Z^\prime$, this contribution dominates the others. 
Hence, we can use semi-Compton scattering cross sections as an appropriate approximation. Finally, we estimate the muon coalescence by taking the on-shell muon decay for $m_{Z^\prime} \geq 2 m_{\mu}$ which can lead to $Z^\prime$ production. Hence, the production rate can be calculated as 
\begin{align}
\Gamma_{\mu \mu \to Z'} (\omega, r)  &= \frac{g^2_{Z' \mu\mu} m^2_{Z'}}{6 \omega (e^{\omega/T} -1)} \sqrt{1-\frac{4 m^2_\mu}{m^2_{Z'}}} \left( 1+\frac{2 m^2_\mu}{m^2_{Z'}} \right)~,
\end{align}
where $g_{Z^\prime \nu\nu/\mu\mu}^2= g_X^{2} (C_V^2+C_A^2)$ involving the information of the U(1)$_X$ charges of left- and right-handed muons and left-handed neutrinos from Tables~\ref{tab1} and \ref{tab2}, respectively. 
Hence, the vector and axial-vector couplings $(C_{V,A})$ for the leptons can be found in Table~\ref{tab-3}. 
The temperature profile $T(r)$, muon number density profile $n_\mu(r)$ as well as the suppression factor due to muon degeneracy $F_{\rm deg} (r)$ are determined from simulations in Ref.~\cite{Bollig:2020xdr}. 
There are two supernova models SFHo18.8 and SFHo20.0 which can be used for providing a conservative and an optimistic constraint, respectively. 
We show the SFHo20.0 profile in Fig.~\ref{Supernova1} for different $x_H^{}$ and that for SFHo18.8 is not shown for simplicity as it is less constrained. 
However, in the calculations of the bounds on $g_X^{}$ for different $m_{Z^\prime}^{}$, both bounds are discussed. 
In this analysis, we ignore the subdominant rate of pair neutrino annihilation into $Z^\prime Z^\prime$ due to two powers of U(1)$_X$ coupling suppression compared to the semi-Compton scattering and bremsstrahlung from muons, respectively, following Ref.~\cite{Croon:2020lrf}.
For the case with $x_H^{}=-2$ (upper-left panel of Fig.~\ref{Supernova1}), the coupling between the neutrino and $Z'$ is turned off, and the luminosity is dominated by the semi-Compton scattering.
Therefore, the luminosity is not sensitive to $m_{Z'}$ when $m_{Z'} \lesssim 2 m_{\mu}$. 
While as $m_{Z'} \gtrsim 2 m_{\mu}$, the muon-pair coalescence dominates the production, and smaller coupling can be reached. 
For other values of $x_H^{}$, the lower bounds on $g_{X}^{}$ is mainly determined by the semi-Compton scattering when $m_{Z'}^{} \lesssim 10$ MeV and by the neutrino- and muon-pair coalescence when $m_{Z'}^{} \gtrsim 10$ MeV. 
Note that although the productions of neutrino- (muon-) pair coalescence can be enhanced by large $m_{Z'}^{}$, the lower bounds do not become more stringent for heavier $Z'$ in the region $m_{Z'}^{} \gtrsim 100$ MeV due to the Boltzmann suppression. 
The rescattering and reabsorption effects lead to the dramatic reduction in luminosity at large $g_{X}^{}$, giving rise to the upper bounds on the coupling. 
The Boltzmann suppression renders weaker upper limit for heavier $Z'$. 
We mention that following the arguments given in Refs.~\cite{Bollig:2017lki,Croon:2020lrf}, we do not consider the effect of electrons because electron semi-Compton rate is suppressed by Pauli blocking.~\footnote{
In this analysis, we consider the effect of $Z^\prime$ for simplicity.
However, the effect of plasmon could be considered which will leave the results almost the same according to Ref.~\cite{Caputo:2021rux}. 
Here, we restrict ourselves to the radial movement of the bosons whereas the directions other than radial ones which will reduce the rate appearing in opacity. 
Consideration of the other directions will weaken the upper limit of the cross sections by one order of magnitude. 
It can be simply scaled affecting the upper limit on $g_X^{}$ roughly by a factor of 3.
However, the lower limit will remain the same as Ref.~\cite{Caputo:2022rca}.
}
\begin{figure}
\begin{center}
\includegraphics[width=0.93\textwidth]{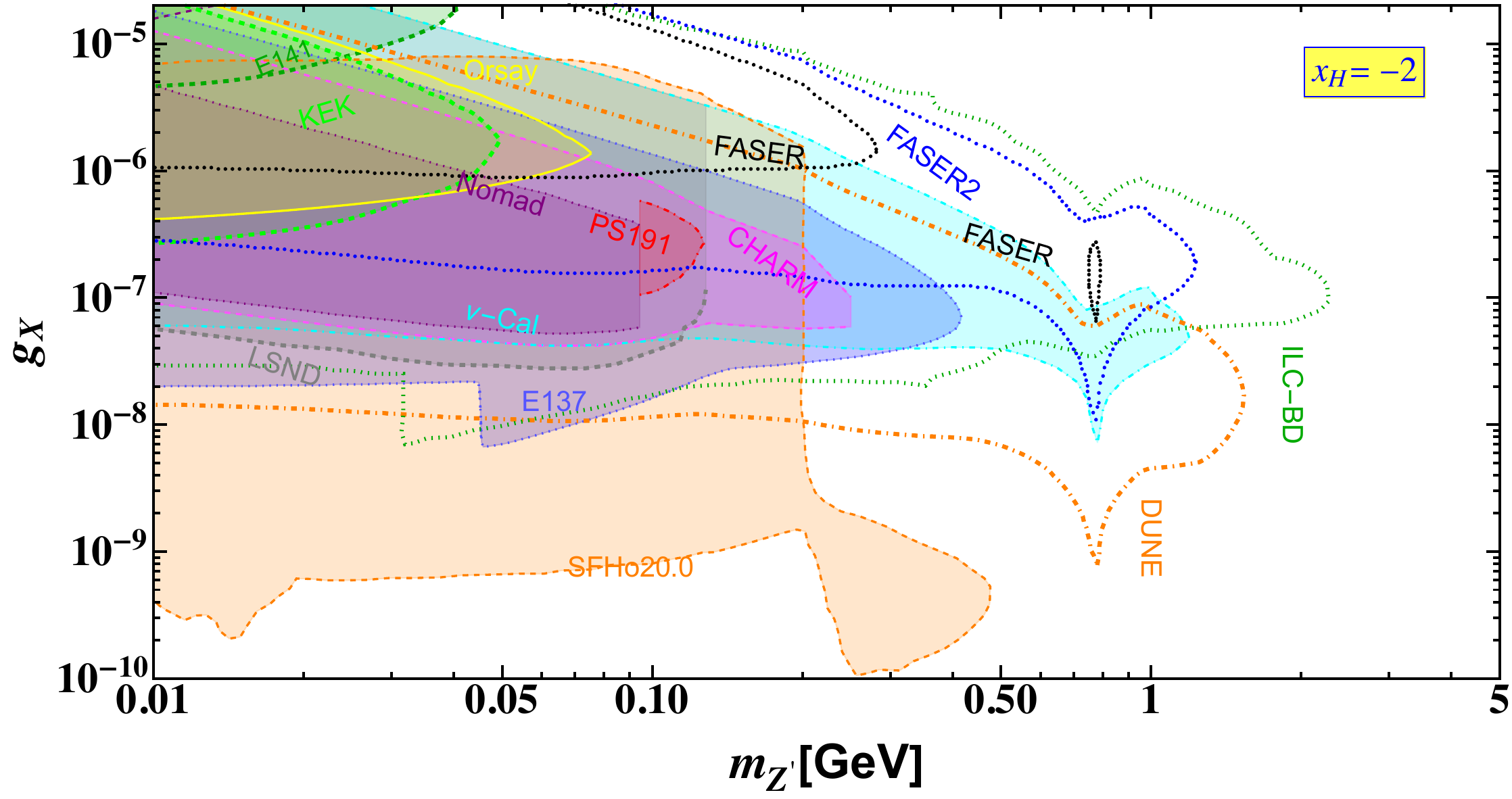}
\includegraphics[width=0.93\textwidth]{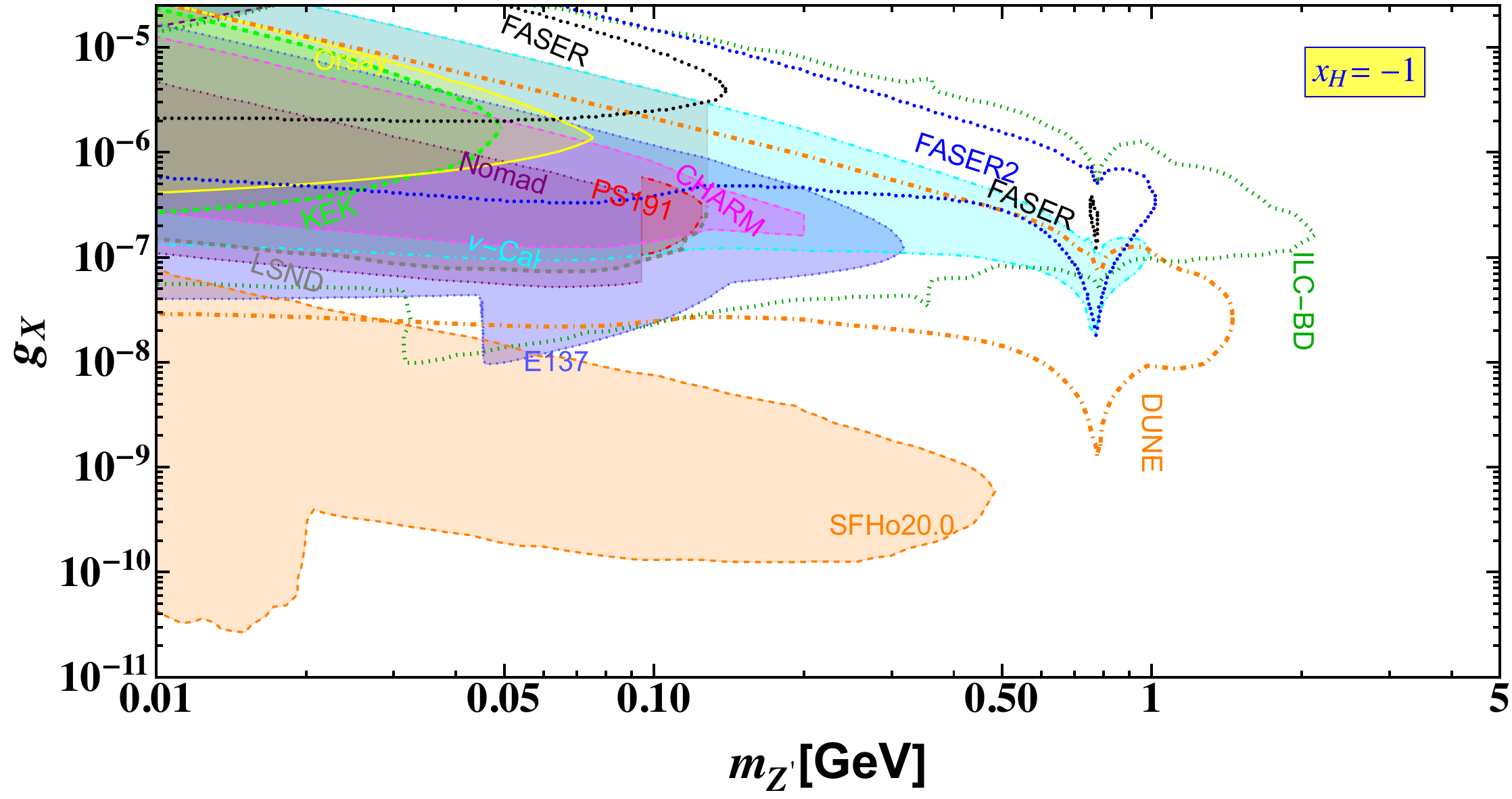}\\
\includegraphics[width=0.93\textwidth]{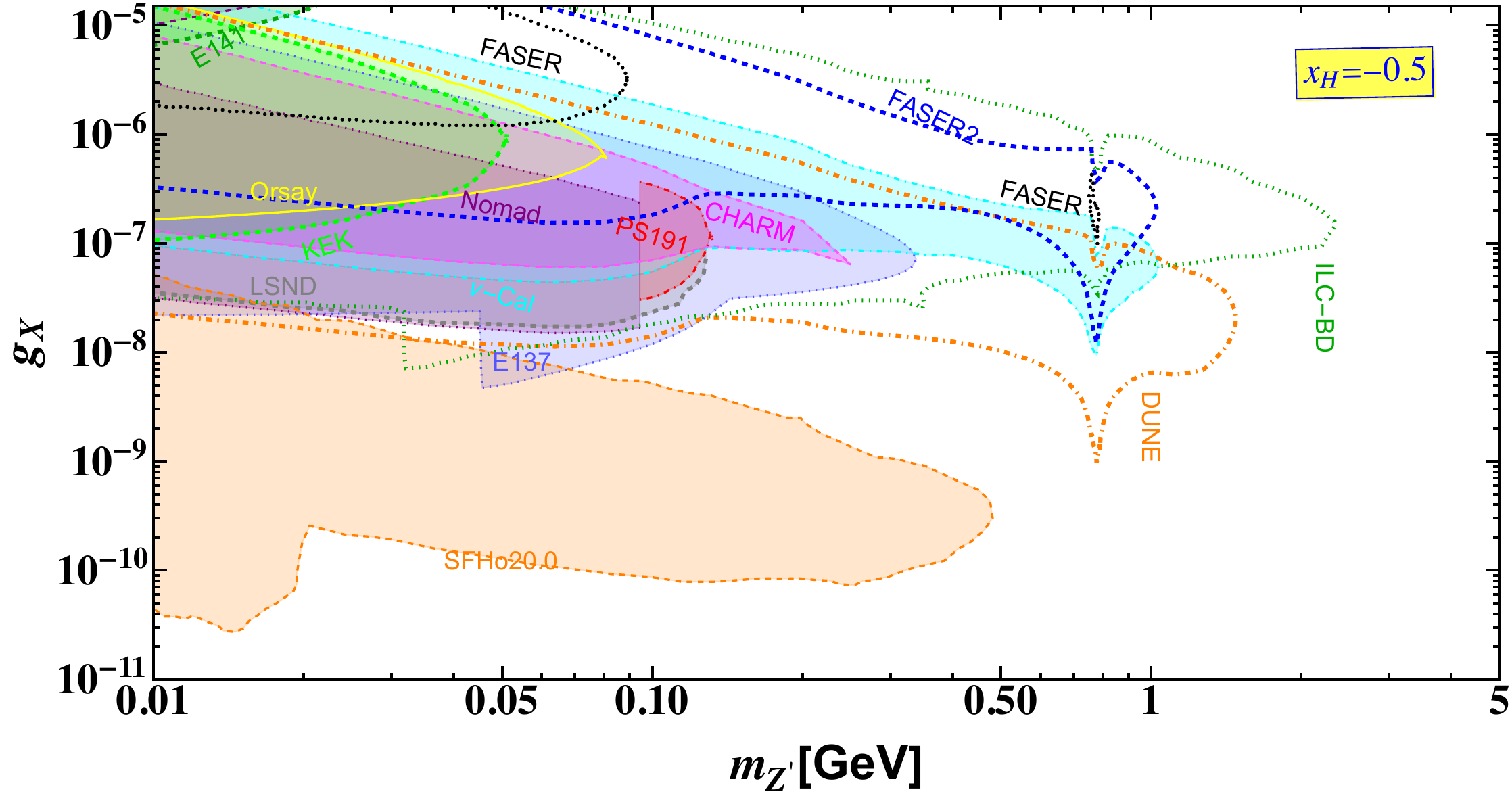}
\caption{Limits on $g_X^{}-m_{Z^\prime}^{}$ plane for $x_H^{} < 0$ and $x_\Phi^{}=1$ considering $10$ MeV $\leq m_{Z^\prime}^{} \leq 5$ GeV showing the regions could be probed by FASER, FASER2, ILC-Beam dump, and DUNE. We compare the parameter space with existing bounds from different beam dump experiments and a cosmological observation of supernova SN1987A(SFH020.0), respectively.}
\label{fig:xH-}
\end{center}
\end{figure}
\begin{figure}[h]
\begin{center}
\includegraphics[width=0.93\textwidth]{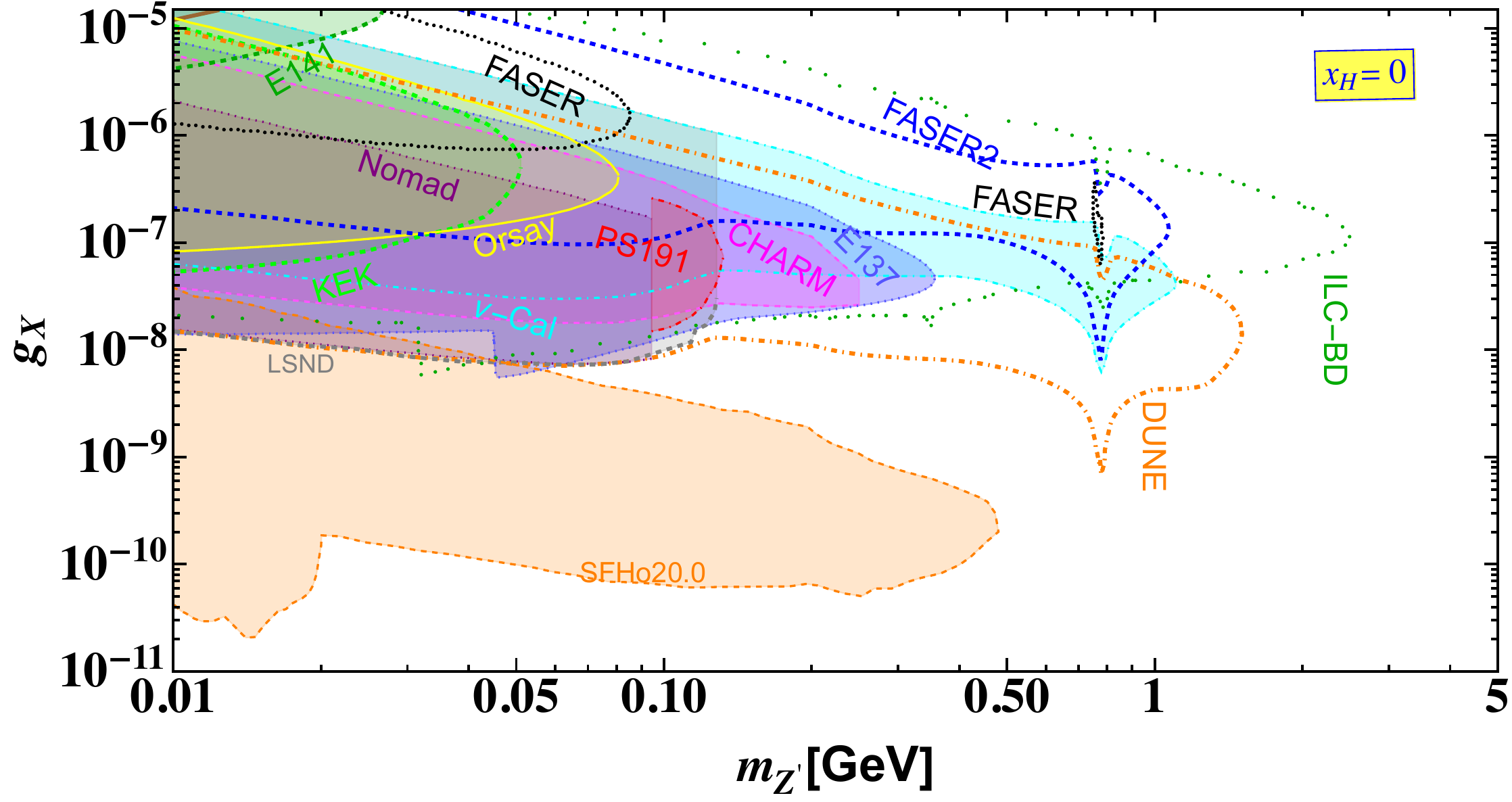}
\caption{Limits on $g_X^{}-m_{Z^\prime}^{}$ plane for $x_H^{}= 0$ and $x_\Phi^{}=1$ considering $10$ MeV $\leq m_{Z^\prime}^{} \leq 5$ GeV showing the regions could be probed by FASER, FASER2, ILC-Beam dump, and DUNE. This is the B$-$L case shown as a reference. We compare the parameter space with existing bounds from different beam dump experiments and a cosmological observation of supernova SN1987A (SFHO20.0), respectively.}
\label{fig:xH0}
\end{center}
\end{figure}
\begin{figure}[h]
\begin{center}
\includegraphics[width=0.93\textwidth]{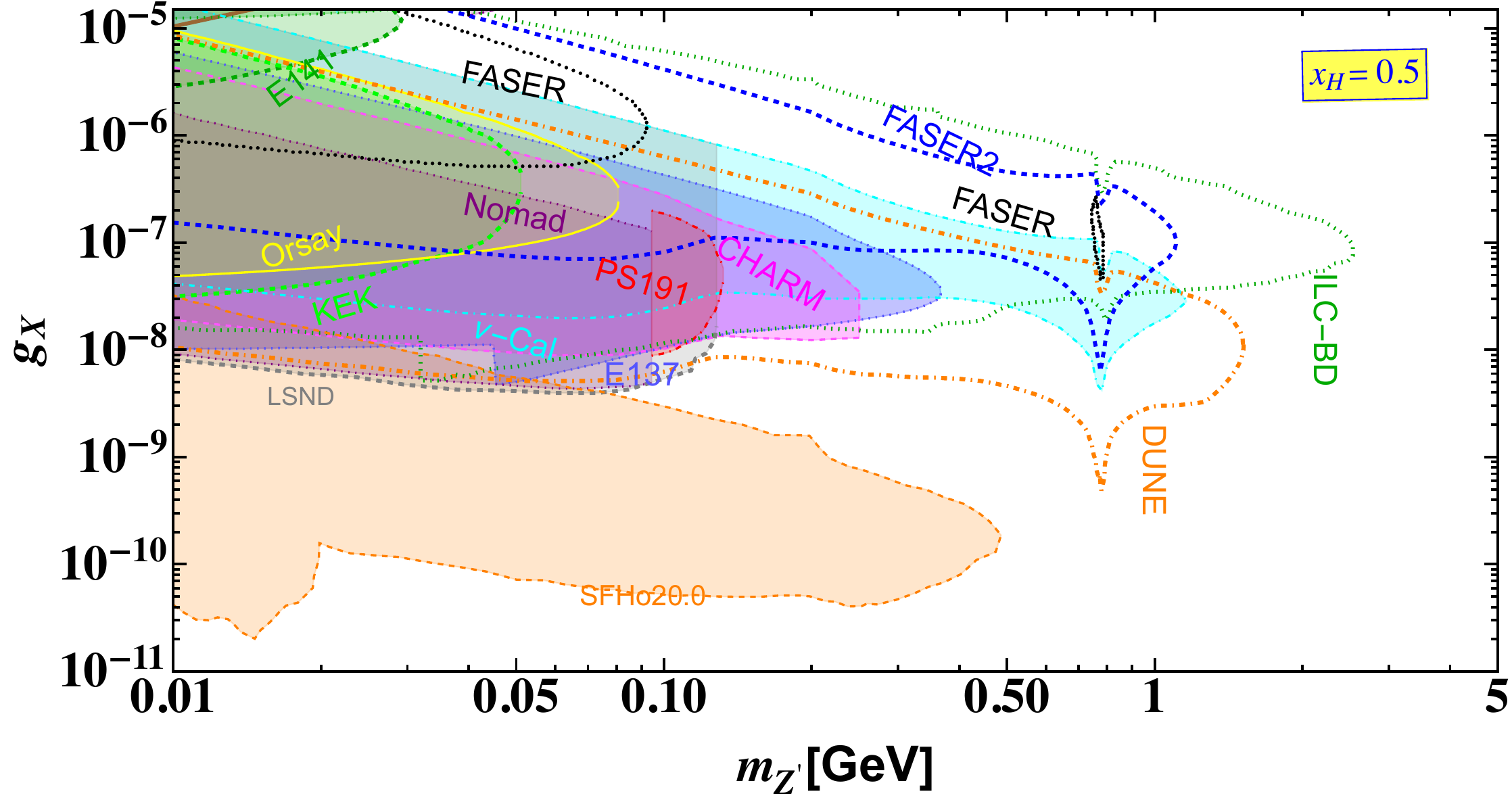}
\includegraphics[width=0.93\textwidth]{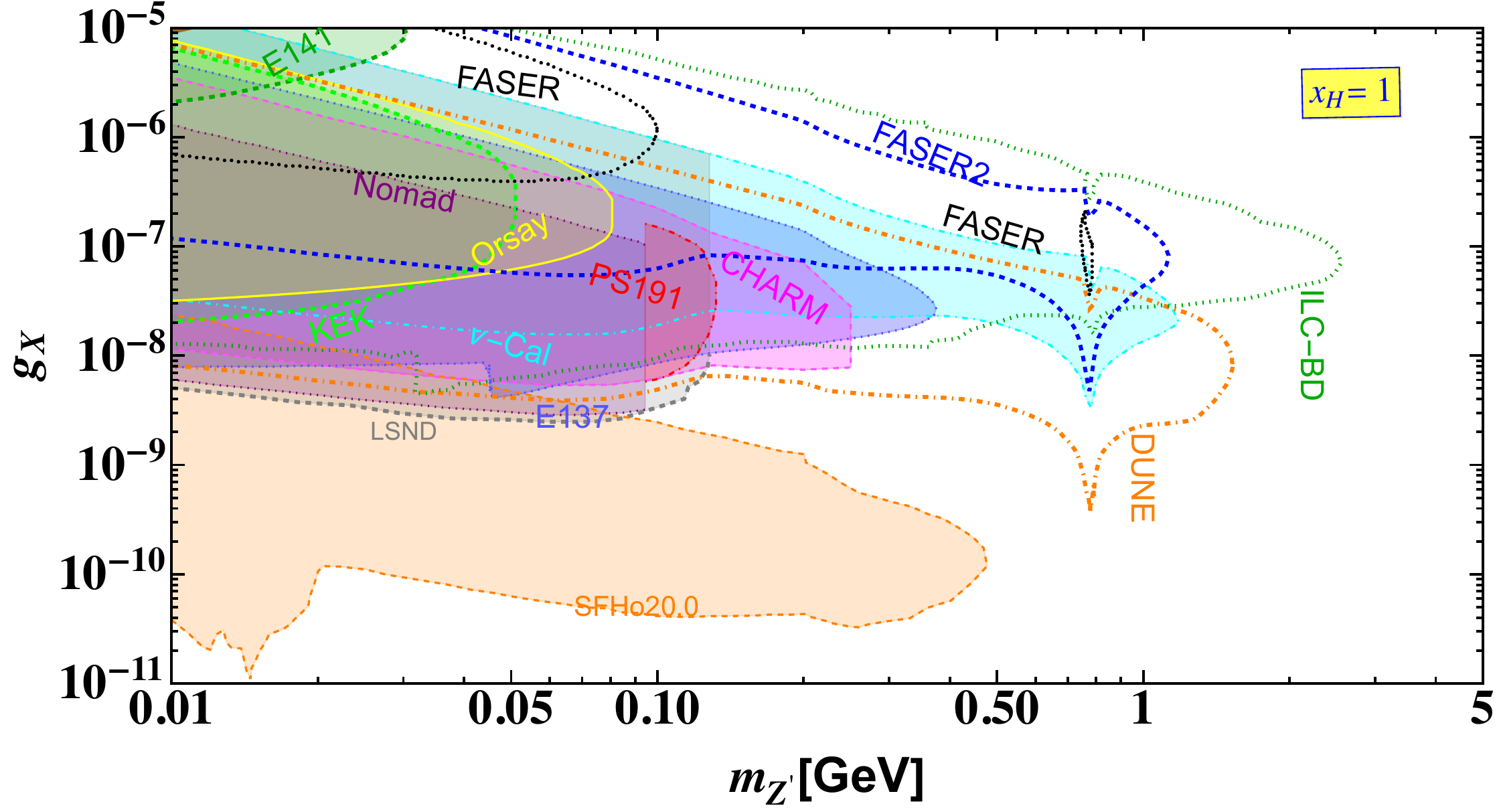}\\
\includegraphics[width=0.93\textwidth]{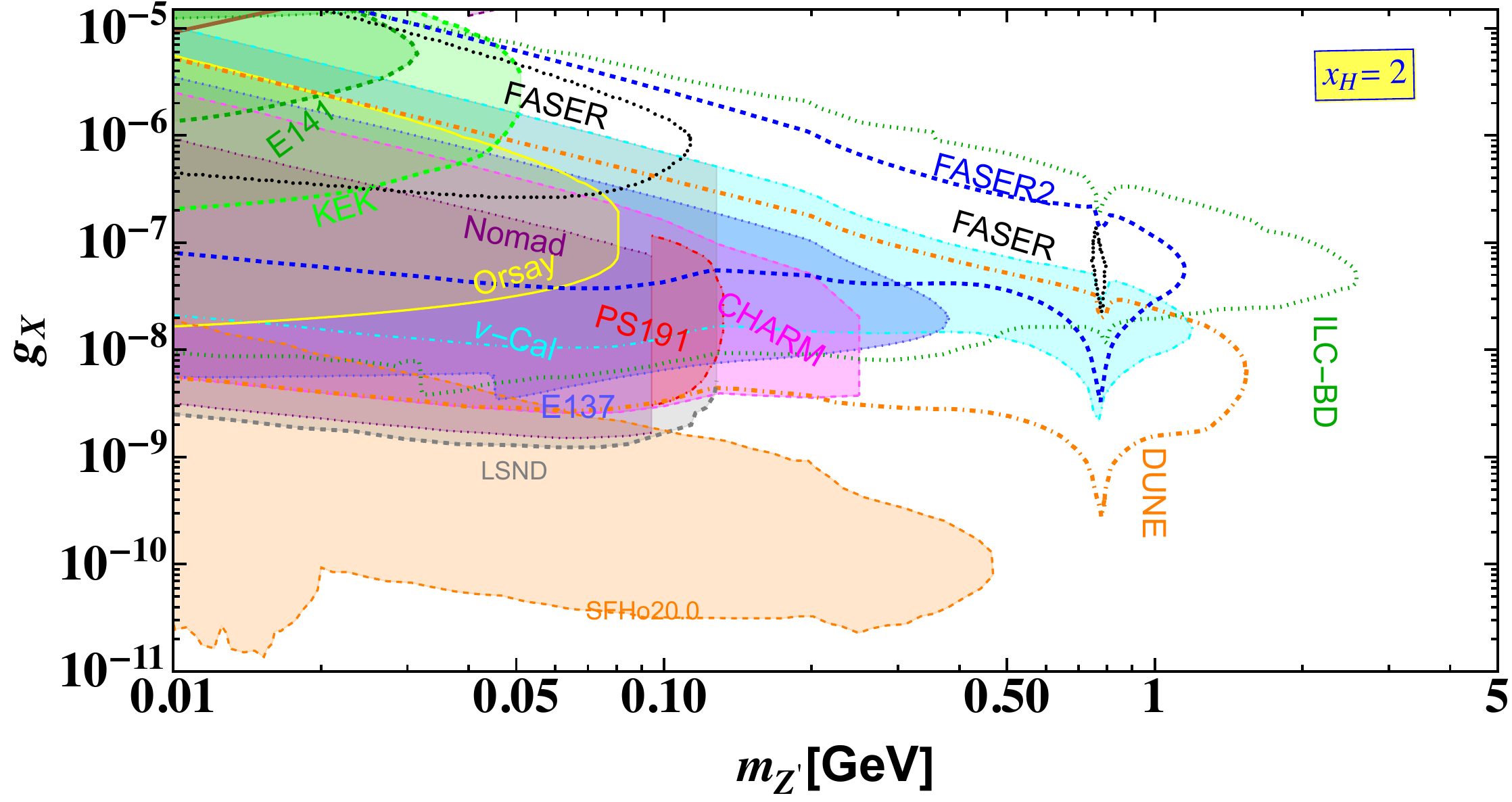}
\caption{Limits on $g_X^{}-m_{Z^\prime}^{}$ plane for $x_H^{}> 0$ and $x_\Phi^{}=1$ considering $10$ MeV $\leq m_{Z^\prime}^{} \leq 5$ GeV showing the regions could be probed by FASER, FASER2, ILC-Beam dump, and DUNE. We compare the parameter space with existing bounds from different beam dump experiments and a cosmological observation of supernova SN1987A (SFH020.0), respectively.}
\label{fig:xH+}
\end{center}
\end{figure}

\section{Results and discussions}
\label{sec:summary}

We calculate the number of the chiral $Z'$ gauge bosons generated by meson decay, bremsstrahlung, and pair annihilation processes at electron, positron, and proton beam dump experiments and FASER(2) and obtain the constraints and future sensitivities of the chiral $Z'$ gauge bosons.
Moreover, we estimate the constraints from SN1987A by discussing energy loss derived from the chiral $Z'$ gauge bosons.

In the calculation of meson production at beam dumps, we calculate the differential production cross section of $\pi^0$ and $\eta$ mesons by the Monte Carlo event generator \texttt{EPOS-LHC}~\cite{Pierog:2013ria} implemented in the \texttt{CRMC} simulation package~\cite{Baus_crmc} for FASER(2) and \texttt{Pythia8}~\cite{Sjostrand:2007gs} for other experiments. 
In those of bremsstrahlung and pair annihilation productions, the track lengths of the electron and positron are calculated by \texttt{EGS5}~\cite{Hirayama:2005zm} code embedded in \texttt{PHITS\,3.23}~\cite{Sato:2018}, and we use the approximate formula shown in Appendix.~\ref{app:track-length}.
The partial decay widths of the chiral $Z'$ gauge boson into lepton pairs are calculated by the analytical formula in Eqs.~\eqref{eq:width-ll}, \eqref{eq:width-nunu}, and \eqref{eq:width-NN}.
On the other hand, in the calculation of that into hadrons for $m_{Z'} < 1.65$ GeV, we assume the VMD and use \texttt{DARKCAST} code~\cite{Ilten:2018crw}.
However, \texttt{DARKCAST} can use for gauge bosons only with vector interactions.~\footnote{
While in final drafting phase, we noticed Ref.~\cite{Baruch:2022esd} studying axial-vector bosons and the new version of \texttt{DARKCAST}.
This version can be used even for the chiral case, and contributions of axial gauge couplings to hadronic decay of the chiral $Z'$ gauge boson can be calculated. 
We compared the branching fractions of the hadronic decay between our result and one by new \texttt{DARKCAST} and confirm that the deviation is at most $10\%$ depending on the choice of $x_H^{}$. 
Therefore, the change derived from the contributions of axial gauge couplings to hadronic decay is small.
In addition to that, we found a disagreement of the formula for the decay width of $Z' \to 2\nu$ in between our paper and the new \texttt{DARKCAST} code and used the modified one in the comparison.
}
Therefore, we calculate the branching fractions into hadrons and electrons in the assumption that only vector interactions exist by \texttt{DARKCAST} and translate them into those in our chiral model as follows:
\begin{align}
    \Gamma(Z' \to {\rm hadrons})_{{\rm chiral}\,Z'} = \Gamma(Z' \to e^+e^-)_{{\rm only\,vector}}~ \frac{{\rm BR}(Z' \to {\rm hadrons})_{\tt DC}}{{\rm BR}(Z' \to e^+e^-)_{\tt DC}}~,
\end{align}
where BR$(Z' \to X)_{\tt DC}$ stands for the branching fraction of the decay $Z' \to X$ calculated by \texttt{DARKCAST}, and
\begin{align}
    \Gamma(Z' \to e^+e^-)_{{\rm only\,vector}} = \frac{m_{Z'}^{} x_{e,V}^{\prime 2} g_X^2}{12 \pi} \left( 1 + \frac{2 m_e^2}{m_{Z'}^2} \right) \sqrt{1 - \frac{4 m_e^2}{m_{Z'}^2}}~,
\end{align}
with $x_{e,V}^\prime = (x_{e,L}^\prime + x_{e,R}^\prime)/2$.
For the calculations of the expected numbers of the signal events, we use \texttt{FORESEE} package~\cite{Kling:2021fwx}. 

We show the limits and sensitivity regions on the $m_{Z^\prime}^{}$-$g_X^{}$ plane for $x_H^{} = -2, -1, -0.5, 0, 0.5, 1$, and $2$ and $x_\Phi^{}=1$ in Figs.~\ref{fig:xH-}, \ref{fig:xH0}, and \ref{fig:xH+}, respectively, as illustrating examples. 
The horizontal and vertical axes are the $Z'$ mass and gauge coupling constant, respectively. 
The boundaries of the bound and sensitivity regions correspond to $N_{95\%}$ events.
Moreover, we also show the constraints from SN1987A, which gives bounds almost in the same mass range of $Z^\prime$ from experiments for long-lived particle search. 
We find the beam dump experiments have sensitivities around $0.01\,{\rm GeV} \lesssim m_{Z'}^{} \lesssim 1\,{\rm GeV}$ and $10^{-8} \lesssim g_X^{} \lesssim 10^{-4}$. 
Particularly, for the lighter $Z'$ mass region than mesons, especially $\pi^0$, the chiral $Z'$ gauge bosons are dominantly produced by rare meson decay. 
Therefore, the bounds from proton beam dump experiments become weaker around $m_{Z'}^{} \simeq 100\,{\rm MeV}$. 
For the E137 and ILC beam dump experiments, there are protrusions around $m_{Z'} \sim \mathcal{O}(100)$\,MeV.
This behavior comes from the enhancement of the $Z'$ production cross section by pair annihilation process. 
For FASER, there are island-shape regions around $m_{Z'}^{} \simeq 0.7$\,GeV.
This behavior results from the enhancement of the electromagnetic form factor of nucleons in Eq.~\eqref{eq:num_p-brem}.

At this point, we mention that, for $m_{Z^\prime}^{} \leq 10$ GeV, big bang nucleosynthesis (BBN) gives a constraint on the U(1)$_{B-L}$ model as $g_{B-L}^{} < \mathcal{O}(10^{-11})$~\cite{Knapen:2017xzo}, and this constraint is stronger than the limits obtained from the beam dump studies.
In the context of BBN from Ref.~\cite{Knapen:2017xzo}, we mention that semi-Compton scattering $(e^- \gamma \to e^- Z^\prime)$ process for $m_{Z^\prime}^{} < 10$ MeV will impose stronger constraints on $g_X^{}-m_{Z^\prime}^{}$ plane. 
Hence, in Figs.~\ref{fig:xH-}, \ref{fig:xH0}, and \ref{fig:xH+}, we show the allowed parameter regions from beam dump experiments for 10 MeV $\leq m_{Z^\prime}^{} \leq 5$ GeV. 
In addition to that stronger bounds from red giants \cite{An:2014twa}, horizontal brunch (HB) starts \cite{Hardy:2016kme}, the Sun \cite{Hardy:2016kme}, fifth force searches \cite{Murata:2014nra}, and neutron scattering \cite{Leeb:1992qf} can be imposed on the gauge coupling for $\mathcal{O}({\rm eV~ scale)} \leq m_{Z^\prime}^{} \leq 10$ MeV in the B$-$L case, and from Ref.~\cite{Knapen:2017xzo}, we find that strongest limit on the gauge coupling could reach at $\mathcal{O}(10^{-15})$.~\footnote{
Estimations of such constraints in the context of the chiral $Z^\prime$ scenarios for $m_{Z^\prime}^{} \leq 10$ MeV are beyond the scope of this current paper. 
However, these studies will be considered in detail in the future to estimate the limits on the gauge coupling from cosmological and astrophysical observations for different $x_H^{}$.
}

Among them, FASER(2), DUNE, and ILC beam dump experiments, which are promising future experiments for long-lived particle search, have higher sensitivities than the past beam dump experiments.
Particularly, the sensitivity of the FASER2 and ILC beam dump can reach $m_{Z'}^{} \sim \mathcal{O}(1)\,{\rm GeV}$ because of its high energy beam and luminosity.
On the other hand, that of DUNE can reach $g_X^{} \sim \mathcal{O}(10^{-9})$ because of its luminosity and long distance between the beam dump and detector.
As mentioned below Eq.~\eqref{eq:qmin}, the production cross section through the proton bremsstrahlung process is enhanced for $m_{Z'}^{} \ll m_p$, and the $Z'$ gauge boson has an axial coupling to the proton, that is, $x_{p,L}^\prime - x_{p,R}^\prime = -x_H^{}/2 \neq 0$. 
Therefore, for the $x_H^{} \neq 0$ cases, DUNE ($\nu$-Cal) gives a stronger sensitivity (bound) than for the $x_H^{} = 0$ (U(1)$_{B-L}$) case.

We obtain excluded region via scaling formulas Eqs.~\eqref{eq:upper} and \eqref{eq:lowerP} from proton beam dump experiments: NOMAD, PS191, LSND, and CHARM.
The constraints are stronger when vector-like $Z'$ couplings to quarks and that to electrons are larger since they are relevant to meson decay to produce $Z'$ and its decay into electron-positron pair. 
Thus, we obtain the strongest constraint for $x_H^{} =2$ case.
Note that NOMAD, PS191, and LSND constraints are relevant up to $Z'$ mass lighter than $\pi^0$ mass since $Z'$ is produced via $\pi^0$ decay at these experiments while slightly larger $Z'$ mass region can be constrained by CHARM since $Z'$ is dominantly produced via $\eta$ decay.
In addition, we obtain excluded regions from electron beam dump experiments: Orsay and KEK applying the scaling formulas Eqs.~\eqref{eq:upper} and \eqref{eq:bremsstrahlung}.
The constraints are stronger when the $Z'$ coupling to electrons is larger since it is related to $Z'$ production cross section and its decay into electron-positron pair. 
Thus we also obtain the strongest constraint for $x_H^{} =2$ case.

The energy loss of SN1987A (SFHo20.0) gives a bound around $m_{Z'}^{} \lesssim 0.5\,{\rm GeV}$ and $10^{-10} \lesssim g_X^{} \lesssim 10^{-7}$.
As have been discussed for Fig.~\ref{Supernova1}, in cases with $x_H^{} \neq -2$, the upper bounds on $g_X^{}$ decrease steadily with increasing $m_{Z'}^{}$ due to Boltzmann suppression. 
And the lower bounds become flat for $m_{Z'}^{} \lesssim 10$ MeV since the $Z'$ production is dominated by semi-Compton scattering. 
For $m_{Z'}^{} \gtrsim 10$ MeV, the pair coalescence dominates, and its production rate is enhanced by large $Z'$ mass. 
The shape of the supernova bound for the $x_H^{}=-2$ case is different from the others because neutrino-pair coalescence is turned off. 
The bound is cut off at $m_{Z'}^{} \sim 2 m_{\mu}$ where the muon decay comes on-shell which prevents the traveling of $Z'$. 
Except for the region with $g_X^{} \sim 10^{-10} - 10^{-9}$ and $m_{Z'}^{} \sim 300$ MeV, the sensitivity is retained for the SFHo20.0 profile, due to the muon-pair coalescence. 
The peak in the around 20 MeV comes from the $Z^\prime$ bremsstrahlung from nucleons $np \to np Z^\prime$ which has been rescaled from Ref.~\cite{Croon:2020lrf}.~\footnote{
For $m_{Z'}^{} \leq 20$ MeV, hadronic activity is dominant over conservative muon profiles. 
A comprehensive analysis of the hadronic modes has been performed in Refs.~\cite{Chang:2016ntp,Rrapaj:2015wgs} which is beyond the scope of this article at this stage.}

\section{Conclusions}
\label{sec:conc}

We have discussed the bounds and sensitivities of the chiral $Z'$ gauge boson by electron, positron, and proton beam dump experiments, FASER(2), and SN1987A.
We take account of rare meson decay and bremsstrahlung processes for all kinds of beam dump experiments and pair annihilation one for electron and positron beam dump experiments.
All the formulae of the $Z'$ production cross section are shown in Sec.~\ref{sec:calc}.
The couplings of $Z'$ to the fermions are generally chiral and depend on the U(1)$_X$ charges of the SM Higgs doublet and U(1)$_X^{}$-breaking singlet scalars.
Therefore, our results in Figs.~\ref{fig:xH-}, \ref{fig:xH0}, and \ref{fig:xH+} are different from each other.

We found that the chiral $Z'$ gauge boson can be explored by experiments for long-lived particle search and SN1987A in broad parameter region.
At this moment, we comment that for the lighter $Z'$ mass than 0.01 GeV, the bound from BBN is the most stringent and could reach $g_X^{} \sim \mathcal{O}(10^{-11})$ specially in the B-L case \cite{Knapen:2017xzo}. 
Such a light $Z^\prime$ is beyond the scope of this paper; however, it will be considered in a future study. For the small coupling region as $10^{-11} \lesssim g_X^{} \lesssim 10^{-7}$, the discussion of the energy loss in SN1987A give a strong bound when 10 MeV $\leq m_{Z^\prime}^{} \leq 0.5$ GeV for different $x_H^{}$, and this, however, has a uncertainty derived from the supernova models.
For the $x_H^{} = -2$ case, the behavior of the bound from SN1987A is completely different from the other cases.
This is because the left-handed lepton doublet has no coupling with $Z^\prime$ for $x_H^{}=-2$. The result is obtained only from the interaction between the right-handed charged leptons and $Z^\prime$ where semi-Compton scattering and muon-pair coalescence will contribute. For other choices of $x_H$, both left- and right-handed leptons will contribute.
We noticed that the beam dump experiments could give bounds in the relatively large coupling ($10^{-8} \lesssim g_X^{} \lesssim 10^{-4}$) and broad mass ($\mathcal{O}(1)\,{\rm MeV} \lesssim m_{Z'}^{} \lesssim \mathcal{O}(1)\,{\rm GeV}$) region.
Particularly, FASER(2), DUNE, and ILC beam dump experiments, which are promising experiments in the future, have higher sensitivities than the past experiments.
We assume 10-years run for the calculations of the future beam dump experiments, and $150\,{\rm fb}^{-1} (3\,{\rm ab}^{-1})$ for LHC run 3 (high-luminosity LHC).
We found that the sensitivity of DUNE reaches $g_X^{} \sim \mathcal{O}(10^{-9})$, and that of ILC beam dump experiment does $m_{Z'}^{} \sim \mathcal{O}(1)$\,GeV.
Therefore, these future experiments are important for the exploration of the chiral $Z'$ gauge boson. In this paper, we consider FASER(2), DUNE, and ILC beam dump experiments as future experiments.
Besides these, there are other proposed experiments, for example, HPS~\cite{Moreno:2013mja}, SeaQuest~\cite{Gardner:2015wea}, MATHUSLA~\cite{Chou:2016lxi}, CODEX-b~\cite{Gligorov:2017nwh}, and AL3X~\cite{Gligorov:2018vkc}, and they may have sensitivities to chiral $Z^\prime$ gauge bosons.
We leave studies of constraints from them for another occasion.

Finally, we note that if the number of the expected events is enough, there is a possibility that the information about the helicity of daughter visible particles and the ratio of the number of detected leptons and hadrons help the discrimination of models.
We leave this study for future work.

\section*{Acknowledgments}
We thank Tomoko Ariga and Felix Kling for useful advice. We thank Edoardo Vitagliano for useful communications. This work is supported by JSPS KAKENHI Grant No.\,18H01210, No.\,21K20365 [KA], No.\,19K03860, No.\,19K03865, and No.\,21H00060 [OS], by MEXT KAKENHI Grant No.\,18H05543 [KA], by the Fundamental Research Funds for the Central Universities [TN, JL], and by the National Natural Science Foundation of China (NNSFC) under grant number 11905149 [JL].

\appendix
\section{TRACK LENGTH}
\label{app:track-length}

The track length $l_i$ of a particle $i$ is a useful variable to estimate how many particles are produced in electromagnetic showers.
This variable depends on particle species, beam energy, and material of the beam dump.
For example, in the case of the electron beam, the track length of electrons consists of the contributions from not only electromagnetic shower, but also beam electrons. 
For removing the dependence on the material detail, we introduce the dimensionless normalized track length defined by
\begin{align}
\label{eq:track-length}
    \hat{l}_i \equiv \frac{\rho}{X_0} l_i~,
\end{align}
where $\rho$ and $X_0$ stand for the density and radiation length of the beam dump material, respectively.

\begin{figure}[t]
\centering
\includegraphics[clip=true, scale=0.45]{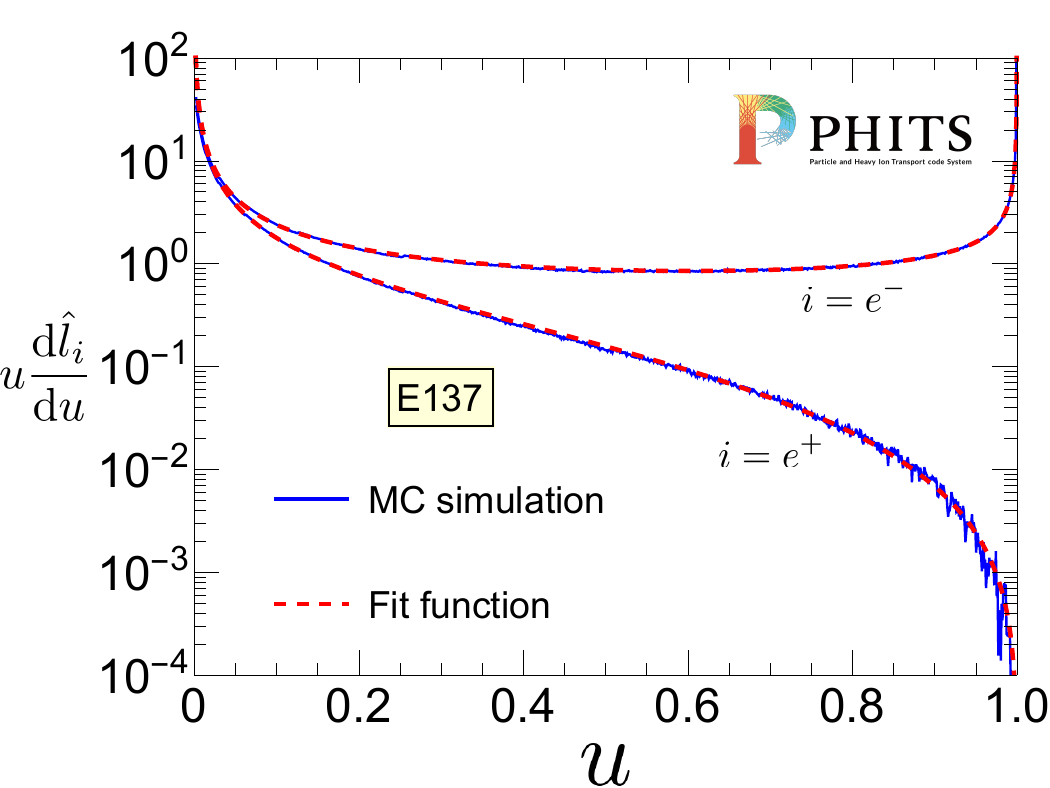}
\includegraphics[clip=true, scale=0.45]{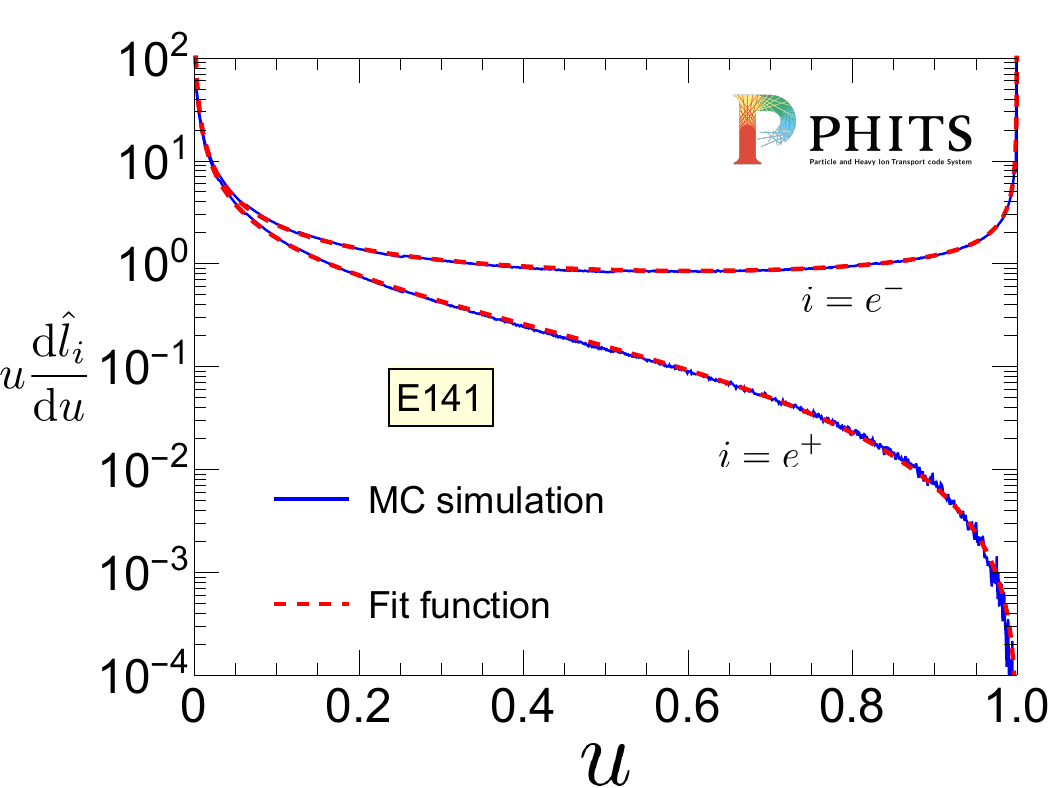}
\caption{
Normalized differential track lengths of electron and positron in the beam dumps of E137 and E141 experiments.
The Monte Carlo simulation has been done by \texttt{PHITS\,3.23}~\cite{Sato:2018}.
}
\label{fig:track-length}
\end{figure}

In this paper, we use \texttt{EGS5}~\cite{Hirayama:2005zm} code embedded in \texttt{PHITS\,3.23}~\cite{Sato:2018}, which is one of calculation code for Monte Carlo simulation, and estimate the track length of electron and positron in beam dump.
In Figs.~\ref{fig:track-length}, we show the results of the Monte Carlo simulation.

As a result of the simulation, we confirm that the fitting formula of the track length in Ref.~\cite{Asai:2021ehn} gives a good agreement with our results calculated by \texttt{PHITS}.
The fitting formula is given by~\cite{Asai:2021ehn}
\begin{align}
     \frac{\dd l_{i}}{\dd E_i} = 
     \frac{X_0}{\rho E_{\rm beam}}\frac{\dd \hat{l}_{i}}{\dd u}~,
\end{align}
where $u = E_i/E_{\rm beam}$ with $E_{\rm beam}$ being the energy of the incident particle, and
\begin{alignat}{2}
   u\frac{\dd \hat l_{e}}{\dd u} &=
   \left( u \frac{\dd \hat l_{e}}{\dd u} \right)_{\rm primary} + \left( u \frac{\dd \hat l_{e}}{\dd u} \right)_{\rm shower} \quad
   && \text{($e^\pm$ from $e^\pm$ beam)}~, \\
   u \frac{\dd \hat l_{e}}{\dd u} &=
   \left( u \frac{\dd \hat l_{e}}{\dd u} \right)_{\rm shower}
   && \text{($e^\mp$ from $e^\pm$ beam)}~,
\end{alignat}
with
\begin{align}
   \left( u \frac{\dd \hat{l}_{e}}{\dd u} \right)_{\rm primary} &=
   0.581 + 0.131 \left( \frac{u}{1-u} \right)^{0.7}~, \\
   \left( u \frac{\dd \hat{l}_{e}}{\dd u} \right)_{\rm shower} &=
   \frac{1-u}{u}(a - b u^2)~,
\end{align}
where $(a, b) = (0.199, 0.155)$ for ILC beam dump, and $(0.199, 0.170)$ for E137 and E141.

\bibliography{bibliography}
\bibliographystyle{utphys}
\end{document}